\documentclass[twocolumn]{aastex631}

\usepackage{amsmath}
\usepackage{bm}
\usepackage{graphicx}

\newcommand{\msol}{$M_\odot$~}
\newcommand{\ron}{$R_{1.4}$~}

\begin{document}

\title{Equation of State Independent Determination on the Radius of a 1.4 \msol Neutron Star Using Mass-Radius Measurements}

\author[0000-0001-6406-1003]{Chun Huang}
\affil{Physics Department and McDonnell Center for the Space Sciences, Washington University in St. Louis; MO, 63130, USA}
\correspondingauthor{Chun Huang}
\email{chun.h@wustl.edu}


\begin{abstract}
Traditional methods for determining the radius of a 1.4 \msol neutron star (\ron) rely on specific equation-of-state (EOS) models that describe various types of dense nuclear matter. This dependence on EOS models can introduce substantial systematic uncertainties, which may exceed the measurement uncertainties when constraining \ron. In this study, we explore a novel approach to constraining \ron using data from Neutron Star Interior Composition Explorer observations of PSR J0030+0451 (J0030) and PSR J0437-4715 (J0437). However, this work presents a more data-driven analysis framework, substantially decreasing the need for EOS assumptions. By analyzing the mass-radius measurements of these two neutron stars, we infer \ron using statistical methods based mostly on observational data. We examine various hotspot configurations for J0030, along with new J0437 observations, and their effects on the inferred radius. Our results are consistent with X-ray timing, gravitational-wave, and nuclear physics constraints, while avoiding EOS-related biases. The same method has also been applied to a simulated mass-radius data set, based on our knowledge of future X-ray telescopes, demonstrating the model's ability to recover the injected \ron value in certain cases. This method provides a data-driven pathway for extracting neutron star properties and offers a new approach for future observational efforts in neutron star astrophysics.
\end{abstract}



\section{Introduction} \label{sec:intro}
The mass of a typical neutron star ranges from 1 to several \msol. Investigating the upper mass limit of neutron stars can provide strong constraints on the equation of state (EOS) for dense matter, as demonstrated by X-ray observations of PSR J0740+6620 \cite{Miller21,Riley21,Salmi2022,salmi2024,2024ApJ...974..295D} and radio observations \cite{Fonseca21}. However, it is also important  to precisely compute or measure the radius of a neutron star with a mass around 1.4 \msol (\ron), as this value corresponds to the pressure of neutron-rich matter at approximately twice nuclear saturation density \cite{Lattimer2001,Lattimer2013,Drischler21,lim24}. Accurate determination of \ron can provide critical insights into the properties of medium-mass neutron stars. Moreover, \ron is closely linked to neutron skin thickness measurements obtained from nuclear experiments, such as C-REX \cite{crex} and P-REX \cite{prex} and the Gravitational wave constraint \cite{Abbott_2017,Abbott_2018}. Recent discrepancies between the results of C-REX and P-REX highlight the importance of independent constraints on \ron from direct astrophysical constraint, which could help resolve the tension between the C-REX and P-REX data, see eg. \cite{Reinhard13,Reinhard21,Reed_2021,kumar23}.

Typically, modeling the radius of a 1.4 \msol neutron star from nuclear or astrophysical constraints require selecting a specific EOS model (eg. \cite{Eemeli,kumar23,Rutherford_2024}) Bayesian inference techniques are then applied to infer the posterior distribution of the EOS parameters, from which the distribution of \ron is computed. Previous studies have explored various EOS models constraining from astrophysical observations or nuclear experiments, including some metamodels such as polytrope (PP) and speed of sound (CS) and physics-based approaches such as relativistic mean field (RMF) theory etc. (eg. \cite{Raaijmakers_2019,Raaijmakers_2020,Raaijmakers_2021,Huang:2023grj,Huang:2024rvj,Rutherford_2024})

Recent advancements in observational techniques, however, have made it possible to achieve considerably precise mass-radius measurements via detailed pulse profile modeling on X-ray observation data. For instance, NASA’s Neutron Star Interior Composition Explorer (NICER) \cite{gendreau2016neutron} can simultaneously model both the mass and radius of a neutron star through X-ray timing while also determining the star's surface hotspot configuration (See \cite{Miller_2019,Riley_2019,Miller_2021,Riley21,Vinciguerra_2024,salmi2024radiushighmasspulsar,Choudhury_2024,2024ApJ...974..295D}). In addition, gravitational-wave detectors like LIGO have provided valuable constraints on the EOS through observations such as the neutron star merger event GW170817 see \cite{Abbott_2017,Abbott_2018,PhysRevX.9.011001}, which revealed information on the stars’ masses and tidal deformabilities.

In this study, we focus primarily on NICER observations, particularly the recent third observation of the millisecond neutron star PSR J0437-4715 (J0437) \cite{Choudhury_2024}, whose mass has been precisely measured via radio observations to be $1.44 \pm 0.07 M_{\odot}$ \cite{Reardon_2024}. NICER’s first observation target, PSR J0030+0451 (J0030) \cite{Miller_2019, Riley_2019}, initially lacked a mass measurement. Two independent groups working with NICER provided separate mass-radius measurements for this source. \cite{Riley_2019} reported a mass of $1.34_{-0.16}^{+0.15} \, M_{\odot}$, while \cite{Miller_2019} determined a mass of $1.44_{-0.14}^{+0.15} \, M_{\odot}$ , {and they both also provided} constraints on the star’s radius. Subsequent refinements have shown that different hotspot configurations can significantly influence the inferred mass-radius measurement for J0030 (See \cite{Vinciguerra_2024}). which brings a new question of how to correctly model the hotspot configuration. Introducing a more physics-motivated hotspot configuration could be one choice, and this will be studied in \cite{chun24hotspot}.

As discussed earlier, most modeling work of the 1.4 \msol neutron star radius relies on a specific equation of state framework. However, with the recent NICER observations of J0437 and detailed analysis of J0030, both stars close to 1.4 \msol, it is possible statistically to extract \ron information directly from observational data {and} {substantially decrease} the {dependence} of specification on EOS model choices. We must note that the NICER inference results are not entirely EOS independent. For instance, as discussed in \cite{2019ApJ...887L..26B}, the assumption of an oblate neutron star surface in the analyses is based on a preselected set of EOSs. Although this effect is estimated to be minor for NICER sources, which do not rotate too rapidly, it nonetheless introduces a degree of dependence. In this study, we focus on NICER's observations of J0030 and J0437 to derive \ron purely from the data. We compare different mass-radius possibilities for J0030 by exploring various hotspot configuration models.

We consider two cases: first, treating J0030 and J0437 as stars with two different masses, and second, assuming both stars have the same mass of 1.4 \msol but excluding the discussion on the twin star. This approach offers a novel, data-driven method for constraining EOS-related observables, establishing a new framework for using NICER and future telescopes to extract \ron directly from the mass-radius measurements that close to 1.4 \msol.

The Letter is structured as follows: Section 2 introduces the methodology and data used; Section 3 presents the current NICER observation-based inference results under the two different scenarios; In Section 4, we discuss the limitations of current observations and present a future case study to demonstrate the effectiveness of this method; and Section 5 discusses the implications of these methods for current observation and compares them with results from other observations and experiments.
\begin{figure*}
    \centering
    \begin{minipage}{0.32\linewidth}
        \centering
        \includegraphics[width=1\linewidth]{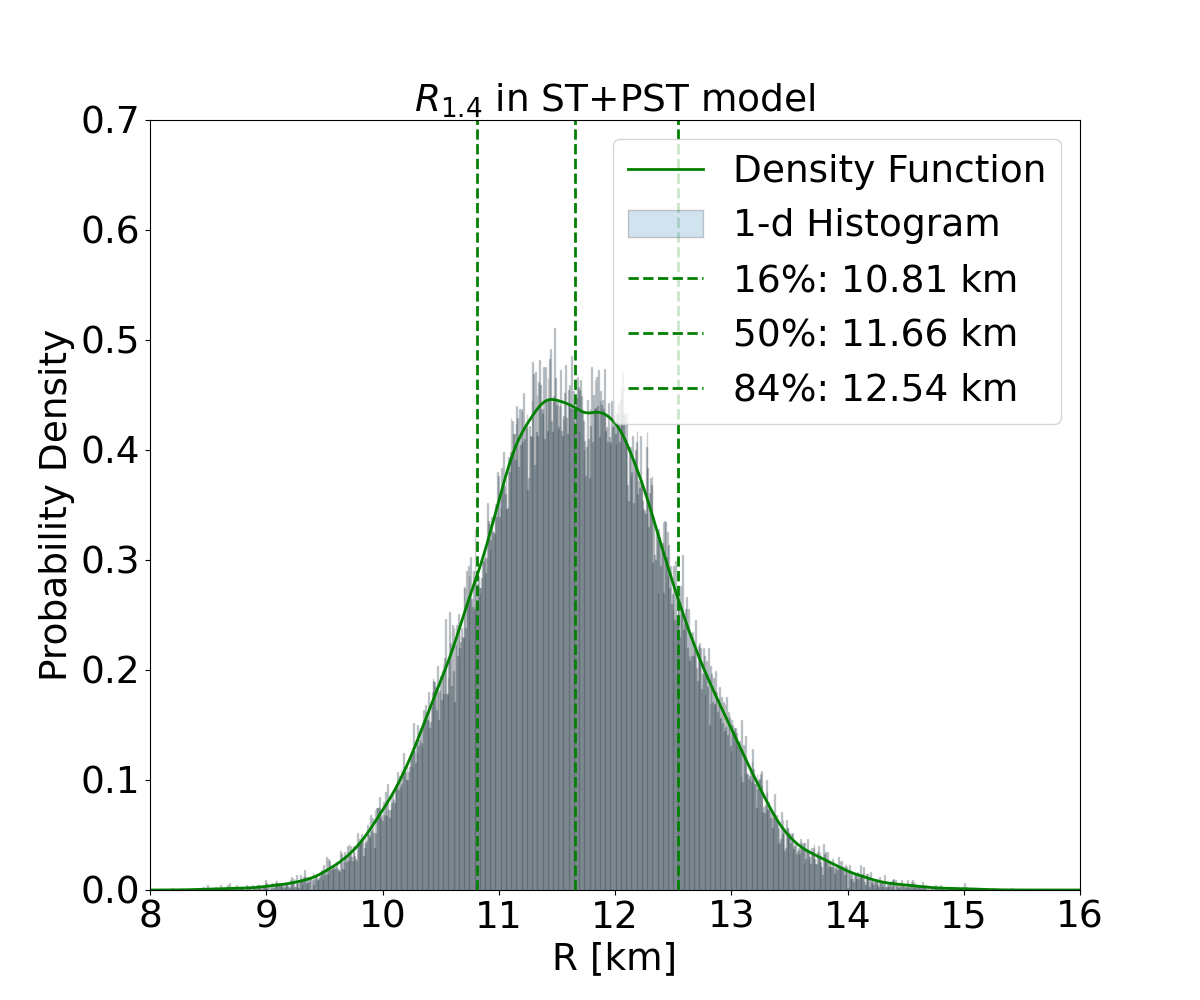}
        \label{stpst}
    \end{minipage}%
    \begin{minipage}{0.32\linewidth}
        \centering
        \includegraphics[width=1\linewidth]{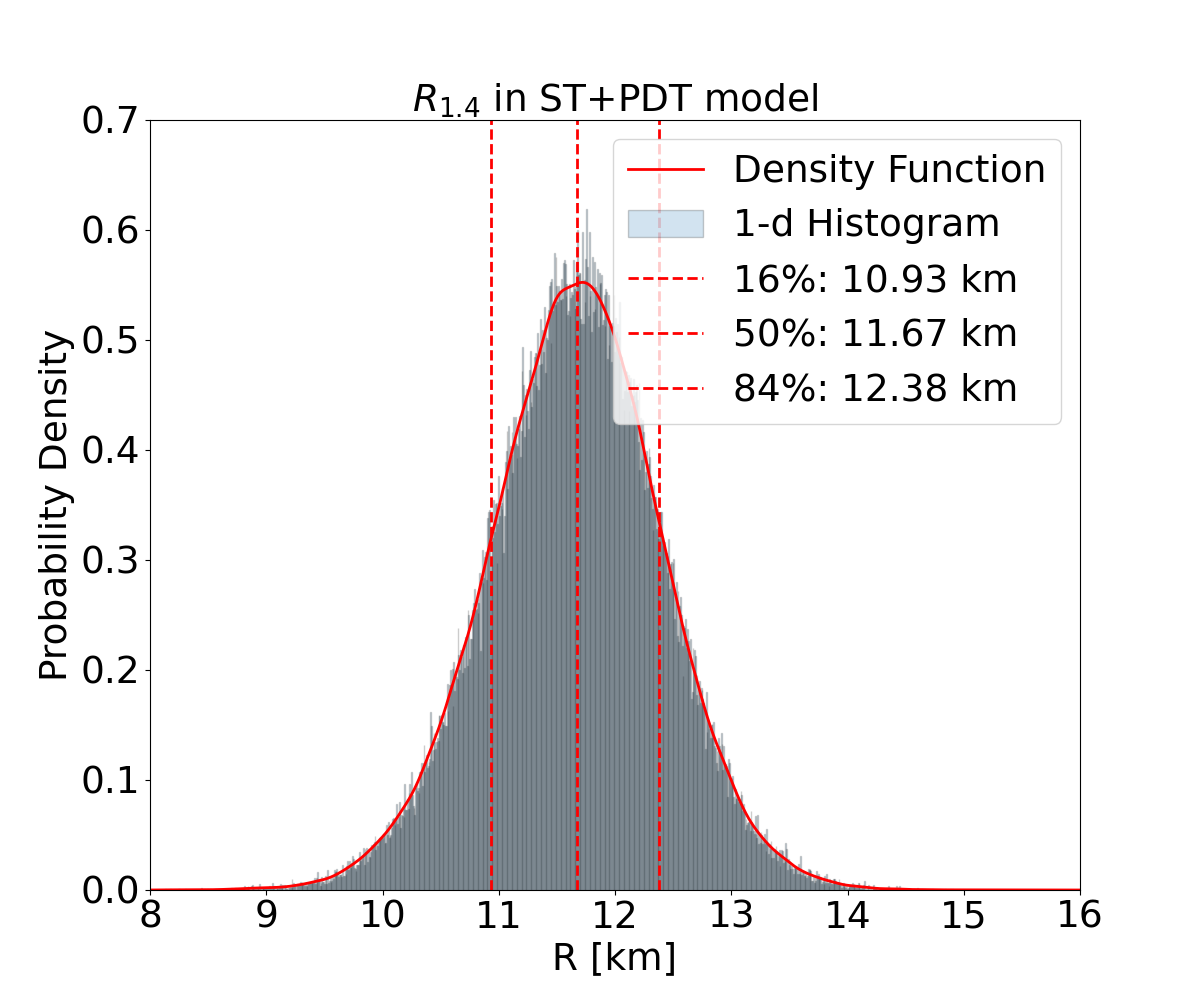}
        \label{stpdt}
    \end{minipage}%
    \begin{minipage}{0.32\linewidth}
        \centering
        \includegraphics[width=1\linewidth]{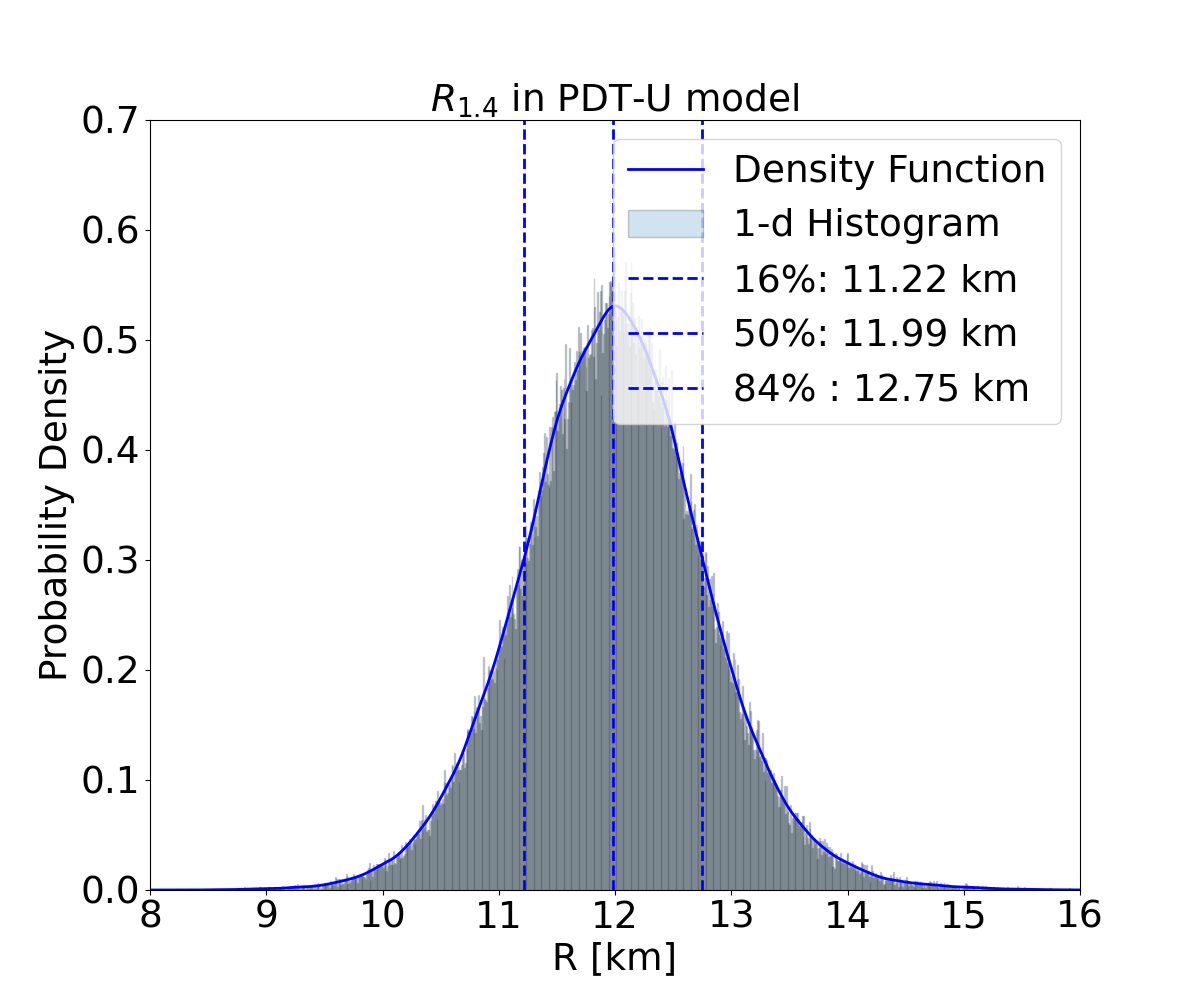}
        \label{pdtu}
    \end{minipage}
    \caption{The 1-D distribution of 1.4~\msol{} star radius \ron{} from joint J0030 and J0437 inference results. With mass-radius measurements of J0030 resulting from (a) ST+PST model, (b) ST+PDT model, and (c) PDT-U model. The dashed lines represent the quantiles of each distribution at 16\%, 50\%, and 84\%. The density function is computed using KDE estimation.}
    \label{fig:Tmap_choices}
\end{figure*}
\section{Methodology and implemented data}
In this section, we will outline our methodology with statistical rigor, clearly defining the assumptions and presenting two distinct scenarios. The data used in this analysis are primarily drawn from NICER observations, with a particular focus on the combined insights from J0030 and J0437 in determining the \ron. This section will emphasize the synergy between these two sources in constraining \ron through observational data, independent of EOS modeling.
\subsection{Inference Methodology}
Two different scenarios will be explored. The first scenario considers the possibility that J0030 and J0437 are distinct stars with different masses and radii. The second scenario examines the case where J0030 and J0437 have the same mass. These two cases complement each other, covering a broad range of possibilities for the actual masses and radii of J0030 and J0437, thereby providing a more comprehensive analysis. In the first scenario, we treat PSR J0030+0451 and PSR J0437$-$4715 as separate observations, each providing mass ($M_i$) and radius ($R_i$) measurements with associated uncertainties. The posterior distributions of mass and radius for each neutron star $i$ (where $i = 1$ for J0030 and $i = 2$ for J0437) are derived from their respective observational data $O_i$. These posterior distributions are converted into continuous probability density functions (PDFs) using kernel density estimation (KDE) in Python:

\begin{equation}
P(M_i, R_i \mid O_i) \approx \mathrm{KDE}(M_i, R_i).
\end{equation}

We then generate $10^6$ random samples of $(M_i, R_i)$ (after excluding the equal-mass ones) for each neutron star from their respective PDFs. Since both neutron stars have mass measurements near 1.4 \msol, we exclude pairs where $M_1 = M_2$, as this would suggest the presence of ``twin stars"---neutron stars with the same mass but different radii. Such a scenario would imply a phase transition in the neutron star equation of state, which is beyond the scope of this analysis (EOS phase transitions {constrained} by the J0437, J0030 and other NICER observations {will be studied in \cite{Huang}}). 

For each remaining sample pair where $M_1 \neq M_2$, we {randomly} select those in which one mass is below 1.4 \msol and the other is above; that is, we select pairs such that $M_1 < 1.4 M_\odot < M_2$ or $M_2 < 1.4 M_\odot < M_1$. In this case, we test the condition \(M_1 \neq M_2\) with a precision of \(10^{-3}\) \msol, which corresponds approximately to the observational precision. If the mass differences of generated mass-radius samples fall below this threshold, we conclude that a strong phase transition is still required to explain this sample (unless the radii are almost equal as well). Consequently, we exclude such cases from our discussion.

We then apply linear interpolation between the selected pairs to estimate the radius $R_{1.4}$ at a mass of $1.4 M_\odot$. The interpolation is expressed as

\begin{equation} R_{1.4} = R_1 + \left( \frac{R_2 - R_1}{M_2 - M_1} \right) \times (1.4 M_\odot - M_1), \end{equation}

This approach allows us to estimate the radius for a neutron star of 1.4 \msol using purely observational data.  It avoids the need to depend on a specific EOS model. However, it still has a few assumptions, like assuming the mass radius relation to be linear like between these two sources. We argue for these two sources that close enough in mass differences, this approximation could be reasonable.
\begin{figure}
	\centering
	\includegraphics[scale=0.6]{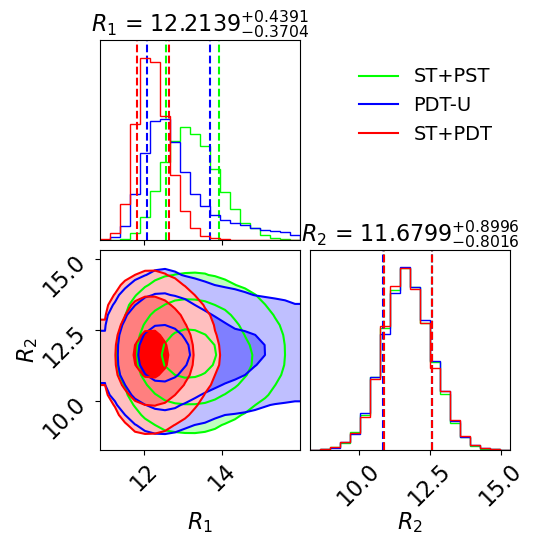}
	\caption{The posterior of $R_1$ and $R_2$ {in scenario 2 }under different hotspot configuration choices of J0030. After applying the condition that all the sources are 1.4 \msol stars. The numbers reported above each 1D marginal distribution refer to the ST+PDT result. Lime green is the ST+PST choice of J0030 plus J0437 resulted posterior, Blue is the one from PDT-U, Red posterior is result from ST+PDT. The contour levels in the corner plot, going from deep to light colors, correspond to the 68\%, 84\% and 98.9\% levels. The dashed line in the 1D corner plots represents the 68\% credible interval, and the title of this plot indicates the median of the distribution as well as the range of the 68\% credible interval. }
	\label{R1R2_posteiror}
\end{figure} 
\begin{table}
\centering

\begin{tabular}{ccc}
\hline \hline
\text{J0030 Hotspot Model Choice}  &       \text{   }&\text{ln(Z)} \\ 
\hline
\text{PDT-U} &       \text{   }    & -1.53    \\
\text{ST+PST}&       \text{   } &  -1.66     \\

\text{ST+PDT} &       \text{   }    & -2.44     \\

\hline 
\hline
\end{tabular}

\caption{This table gives the global log evidence ($\ln Z$), as returned by Ultranest, for the likelihood of J0030 mass-radius measurements from different Hotspot model: PDT-U, ST+PST and ST+PDT, together with J0437 mass-radius measurement, All These Are in scenario 2}
\label{Bayes_evidence}
\end{table}

\begin{figure}
	\centering
	\includegraphics[scale=0.5]{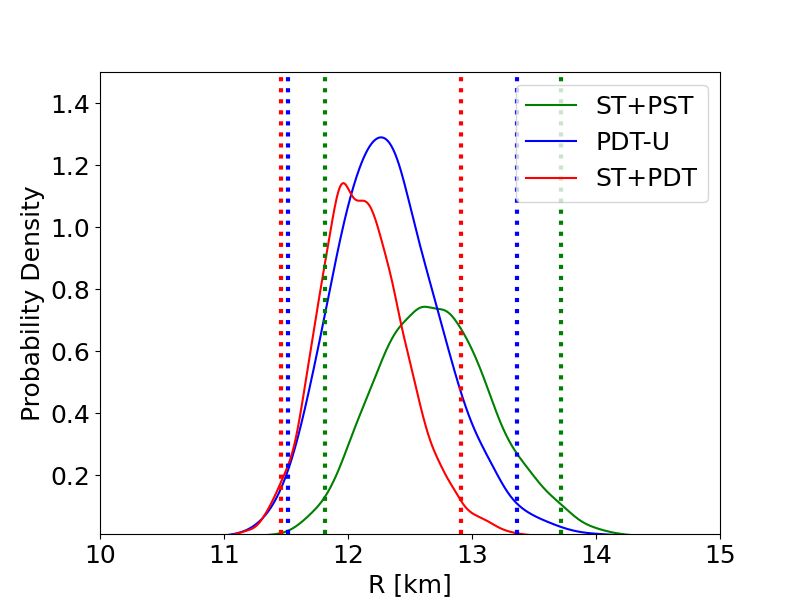}
	\caption{The joint posterior distribution of \ron under different hotspot configuration choices of J0030. After applying the condition that all the sources are 1.4 \msol star. Lime green is the ST+PST choice of J0030 plus J0437 resulted posterior, Blue is the one from PDT-U, Red posterior is result from ST+PDT.  The dashed line with same color as the distribution plots represents the corresponding 95\% credible interval.
	}
	\label{Second_R14_compare}
\end{figure} 
By applying the interpolation method to all suitable sample pairs, we obtain a distribution of $R_{1.4}$ values, representing the possible radii of a neutron star with a mass of $1.4 M_\odot$, inferred solely from the observational data of J0030 and J0437. 

In the second scenario, we consider the case where both J0030 and J0437 have masses of $1.4 M_\odot$. This assumption is particularly relevant for J0437, whose measured mass of $1.4 M_\odot$ lies within the 68\% uncertainty interval. Under this scenario, we employ a Bayesian inference framework.

We define a model with two variables, $R_1$ and $R_2$, representing the radii of J0030 and J0437, respectively, while assuming both stars have a mass of exactly $1.4 M_\odot$. For both radii, we assign uniform prior distributions $Pr(R_i)$ over the range of 6 km to 16 km:
\begin{equation}
Pr(R_i) \sim \mathcal{U}(6\, \text{km}, 16\, \text{km}), \quad i = 1,2.
\end{equation}
This range is sufficiently broad to allow the posterior distributions to converge. Using the mass-radius pairs $(M_i, R_i)$, with $M_i = 1.4 M_\odot$, in conjunction with the previously established KDE function, we compute the likelihood of each star’s radius. The likelihood function for each star is defined as
\begin{equation}
\mathcal{L}_i(R_i) = P(O_i \mid R_i, M_i = 1.4 M_\odot),
\end{equation}
where $O_i$ denotes the observational data for star $i$. Since all mass-radius measurements are, in principle, independent, the total likelihood of this inference is simply the product of the individual likelihoods.

The posterior distributions of $R_1$ and $R_2$ reflect the radius of each neutron star under the assumption of a $1.4 M_\odot$ mass:
\begin{equation}
P(R_i \mid O_i, M_i = 1.4 M_\odot) \propto \mathcal{L}_i(R_i) \cdot Pr(R_i), \quad i = 1,2.
\end{equation}
Since we avoid considering twin star scenarios, we assume the two stars share the same radius, $R_1 = R_2 = R_{1.4}$. Accordingly, we construct a joint posterior distribution for $R_{1.4}$ by taking the product of their posterior distributions while enforcing equality of the radii:
\begin{equation}
\begin{aligned}
P(R_{1.4}) \propto & \; P(R_{1} \mid O_1, M_1 = 1.4 M_\odot) \cdot \\
& P(R_{2} \mid O_2, M_2 = 1.4 M_\odot) \cdot \\
& P(M_1 = 1.4 M_\odot \mid O_1) \cdot \\
& P(M_2 = 1.4 M_\odot \mid O_2)
\end{aligned}
\label{overall}
\end{equation}
Here, $P(M_i = 1.4 M_\odot \mid O_i)$ represents the likelihood of star $i$ having a mass of exactly $1.4 M_\odot$, based on its observational data. The prior $Pr(R_{i})$ is uniform over the range from 6 km to 16 km. This joint posterior distribution represents the inferred radius of a $1.4 M_\odot$ neutron star, based on the combined data from J0030 and J0437 under the assumption that both stars have the same mass and therefore same radius. 

All Bayesian inferences in this study were conducted using the nested-sampling package UltraNest \cite{2021JOSS....6.3001B}. For each inference, we employed 50,000 live points to ensure convergence. All likelihood and computation algorithms are implemented in the CompactObject package \cite{compactobject}, an open-source, full-scope Bayesian inference framework developed by the author, specifically designed for neutron star physics.

By combining the individual posterior distributions and accounting for the likelihood of each star having a mass of exactly $1.4 M_\odot$, we obtain a statistically robust posterior distribution for $R_{1.4}$. This approach integrates the data from both J0030 and J0437, yielding a good estimate of the neutron star radius at $1.4 M_\odot$ under the specified assumptions.

We comment on these two methods as follows. For Scenario 1, due to the limitations of its linear interpolation approach, it is challenging to extend this method to more than two sources, which restricts its constraining power. Scenario 2, on the other hand, can be more readily extended to several observations, particularly those near $1.4 \, M_\odot$. Here, we propose a potential approach to combine Scenario 1 with Scenario 2 to handle more than two sources: if three or more sources with mass measurements close to $1.4 \, M_\odot$ are available, they can be grouped into pairs. For each pair, Scenario 1 can be applied to estimate \ron. This would yield multiple \ron estimations, one from each pair. These estimations can then be combined by computing their joint probability distribution, providing a unified treatment of the sources.

The remaining EOS model dependence of these two methods primarily arises from the assumption of no phase transition in Scenario 1. Specifically, this scenario assumes a smooth linear interpolation between two generated mass-radius points, leading to a smooth mass-radius relation without any discontinuities. Similarly, in Scenario 2, we explicitly exclude the possibility of twin stars, as their inclusion would prevent combining information from two sources into a single distribution of \ron. A detailed discussion of cases involving phase transitions and twin stars is beyond the scope of our analysis here.
 
\begin{table*}
\centering
\begin{tabular}{cccc}
\hline \hline
\text{Scenarios}  &       \text{   J0030 Hotspot Model     }&\text{\ron    }&\text{$\Lambda_{1.4}$} \\ 
\hline
\text{1} &       \text{PDT-U}    & $11.99_{-1.57}^{+1.62} \mathrm{~km}$& $\Lambda_{1.4}=355_{-231}^{+563}$     \\
\text{1}&       \text{ST+PST} &  $11.66_{-1.65}^{+1.77}  \mathrm{~km}$ & $\Lambda_{1.4}=288_{-196}^{+543}$    \\

\text{1} &       \text{ST+PDT}    & $11.67_{-1.50}^{+1.42} \mathrm{~km}$ & $\Lambda_{1.4}=290_{-186}^{+396}$      \\
\text{2} &       \text{PDT-U}    & $12.29_{-0.77}^{+1.02} \mathrm{~km}$ & $\Lambda_{1.4}=429_{-166}^{+354}$    \\
\text{2}&       \text{ST+PST} &  $12.67_{-0.86}^{+1.03} \mathrm{~km}$ & $\Lambda_{1.4}=540_{-221}^{+432}$     \\

\text{2} &       \text{ST+PDT}    & $12.09_{-0.63}^{+0.81} \mathrm{~km}$  & $\Lambda_{1.4}=379_{-126}^{+238}$   \\
\hline 
\hline
\end{tabular}
\caption{This table Summarizes All the predictions for \ron from Different Scenarios, and different J0030 hotspot models and the interpolated Tidal Deformability of derived from Emperical relation $\Lambda\left(R_{1.4}\right)=2.88 \times 10^{-6}\left(R_{1.4} / \mathrm{km}\right)^{7.5}$ in \cite{Eemeli}. {The $R_{1.4}$ values demonstrated as the upper and lower limits correspond to the 95\% credible intervals.}}
\label{R14_table}
\end{table*}
\subsection{Implemented dataset}
In this Letter, we investigate the synergy between PSR J0437$-$4715 and PSR J0030$+$0451. For J0437, we use posterior samples from the recent NICER observations reported by \cite{Choudhury_2024}. The case of J0030 is more intricate: the original analysis by \cite{Riley_2019} suggested a single-temperature circular hotspot in combination with an arch-like hotspot configuration (ST+PST), based on this configuration, they reported a measurement as $M=1.34_{-0.16}^{+0.15} M_{\odot}$ and $R=12.71_{-1.19}^{+1.14} \mathrm{~km}$, based on NICER observation data. \cite{Vinciguerra_2024} refined the analysis by including ST+PDT and PDT-U for the mass-radius measurement of J0030 and updated the ST+PST inference result. In this study, the ST+PST case we use corresponds to the mass-radius measurements reported in \cite{Riley_2019}.

For detailed explanations of these hotspot models, refer to \cite{Vinciguerra_2024}. The ST+PDT and PDT-U configurations are favored in the joint analysis of NICER and XMM-Newton data. Although the PDT-U model has stronger Bayesian evidence, ST-PDT configuration {aligns} more closely with the gamma-ray emission of this source. In the current study, however, we focus only on the mass-radius posteriors informed solely by the NICER data set, since, as discussed in \cite{Vinciguerra_2024,Choudhury_2024}, the joint analysis of the XMM-Newton and NICER data sets either has not been proven robust under different inference settings or has been inferred at very low resolution due to expensive computational requirements. The NICER-only result has been extensively explored with different inference settings. While this may not yet be fully convincing in proving the result’s robustness across different setups, excluding the choices of hotspot geometry models setup, the scatter in results from varying modeling methods is not significant enough based on current observational precision. In this study, these minor discrepancies are well contained within our inference uncertainty region. This approach differs from that of \cite{Rutherford_2024}, where the joint NICER and XMM-Newton results were directly considered in several new scenarios, with subsequent discussion of their implications for neutron star EOS constraints. As a result, these configurations of J0030 hotspots warrant individual discussion to demonstrate how different model setups influence predictions of the radius of a neutron star with a mass of 1.4 $M_{\odot}$ . This approach also enables us to quantify the impact of different hotspot modeling methods and assess the robustness of our results based on varying mass-radius measurements for the same star, as reported from different hotspot models.

\section{Inference Result}
In this section, we discuss the inference of $R_{1.4}$ based on the NICER-only inference data set from the two different methods described above. For each scenario, we compare the results across various hotspot configuration choices for J0030, as different hotspot maps result in distinct mass-radius measurements for J0030, which in turn affect the prediction of $R_{1.4}$.

\subsection{Scenario 1}
In the first scenario, we treat J0030 and J0437 as two distinct neutron stars, each with independent mass and radius measurements. We then perform random sampling to interpolate the radius of a $1.4 M_\odot$ neutron star. The resulting radius distributions vary depending on the choice of hotspot configurations for J0030. 

In Figure \ref{fig:Tmap_choices}, we present the 1D distribution of the $1.4 M_\odot$ star radius based on results from J0030 and J0437 using different hotspot maps. For the ST+PST model, the central value is 11.66 km, with the 16\% quantile at 10.81 km and the 84\% quantile at 12.54 km. This 1D histogram is divided into 1000 bins. With this choice, the fluctuations are already quite small. 

In comparison, the posterior result for $R_{1.4}$ using the ST+PDT configuration, shown in middle panel of Figure \ref{stpdt}, is more concentrated, with a narrower 68\% credible interval. This reflects the fact that the ST+PDT configuration for J0030 produces a mass-radius measurement with smaller uncertainty compared to other measurements, as $M = 1.20_{-0.11}^{+0.14}$\msol, $R = 11.16_{-0.80}^{+0.90} \mathrm{~km}$.

In contrast, in the rightmost panel of Figure \ref{pdtu}, the $R_{1.4}$ distribution derived from the PDT-U model has the largest central value. This makes sense given that the mass-radius measurement for J0030 under the PDT-U configuration is significantly different from other setups and favors a mass range of $M = 1.41_{-0.19}^{+0.20} M_{\odot}$, which is located around the 1.4 solar mass region, and a large radius region of $R = 13.12_{-1.21}^{+1.35} \mathrm{~km}$, quoting the 68\% credible interval here. 


Remarkably, despite the variations in the hotspot configuration models for J0030, the central values of all the radius distributions fall within each other’s 68\% credible intervals. The primary differences across the configurations are reflected in the widths of the 68\% credible intervals.
\begin{figure*}
    \centering
    \begin{minipage}{0.48\linewidth}
        \centering
        \includegraphics[width=1\linewidth]{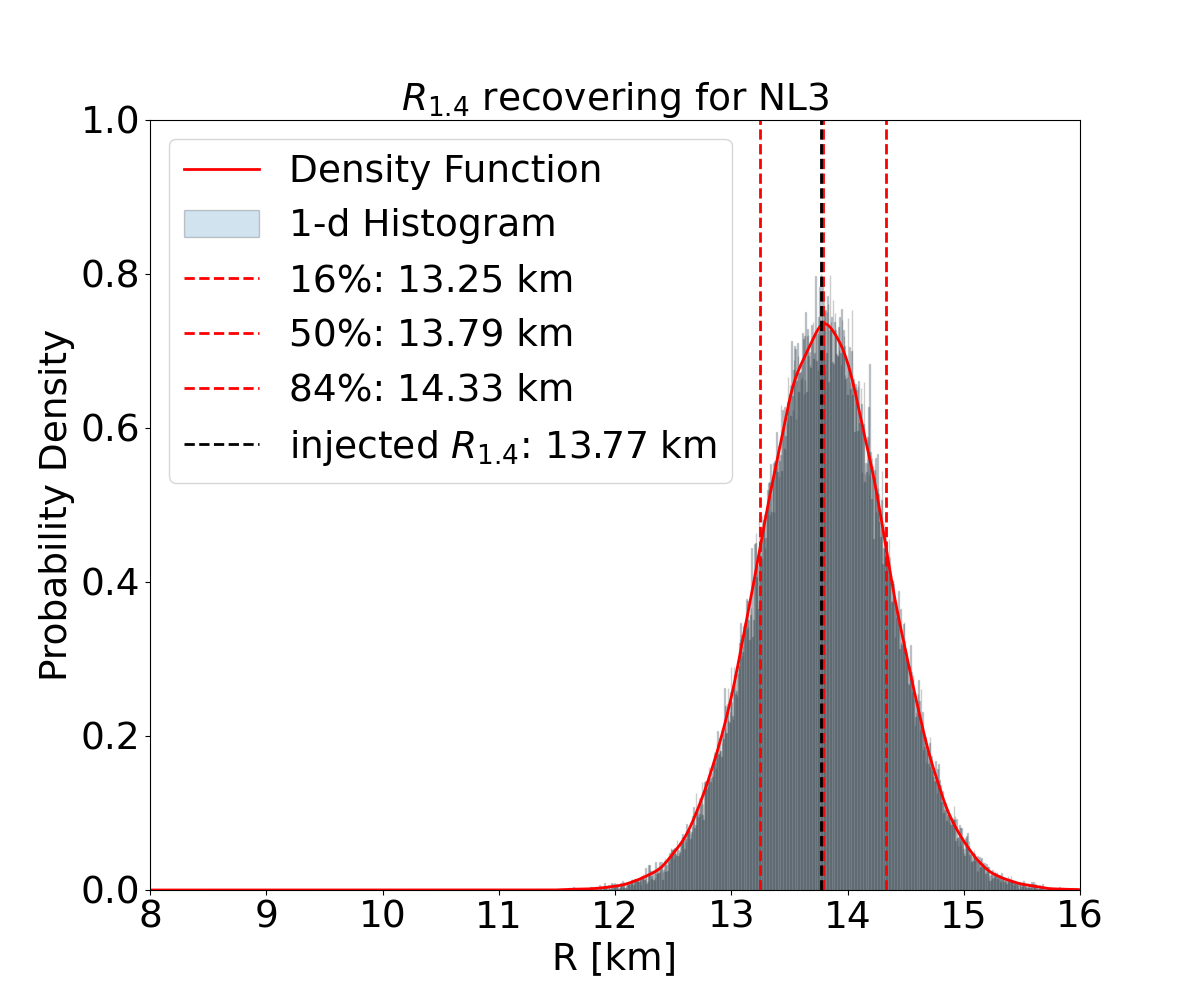}
        \label{NL3}
    \end{minipage}%
    \begin{minipage}{0.48\linewidth}
        \centering
        \includegraphics[width=1\linewidth]{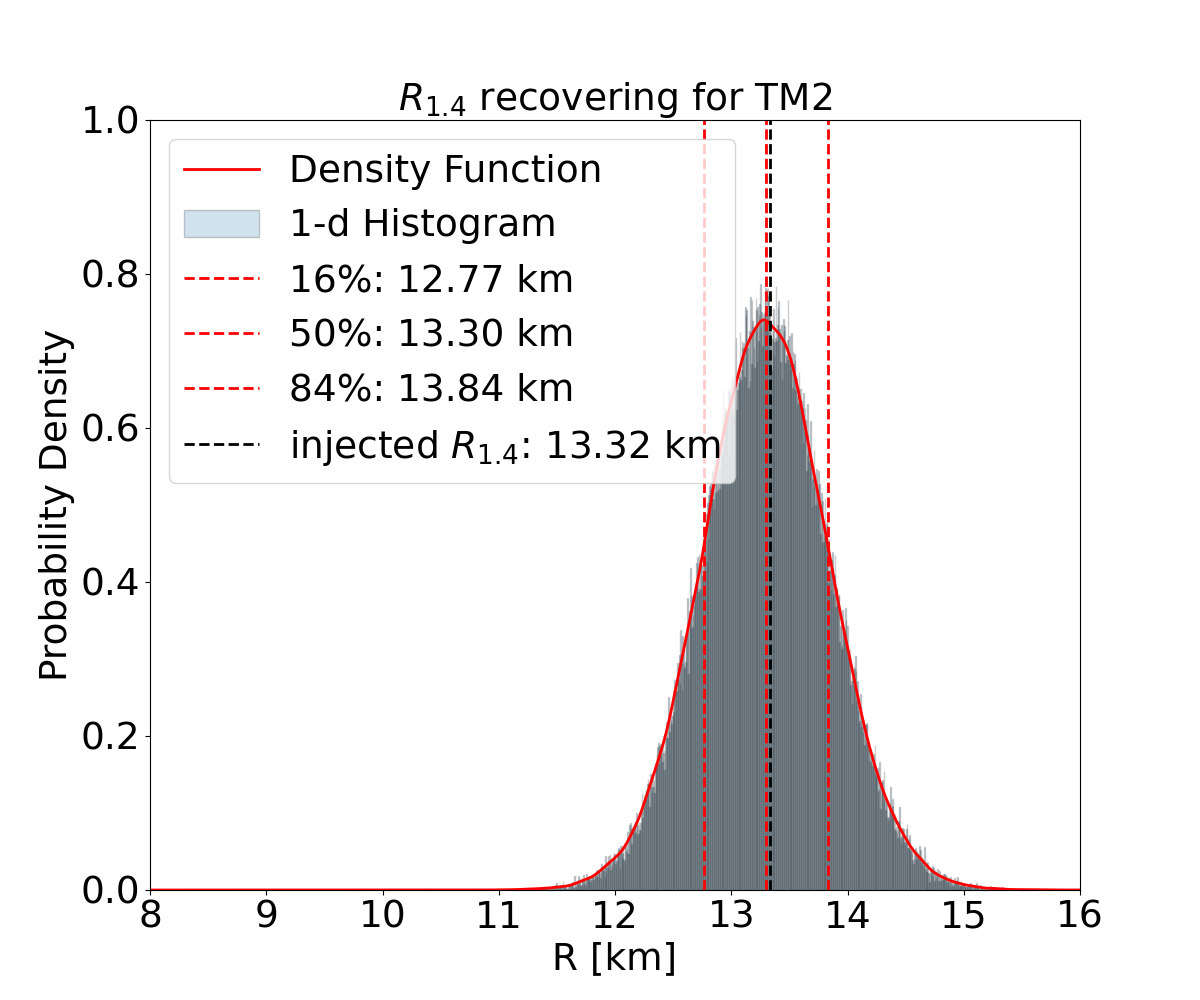}
        \label{TM2}
    \end{minipage}%
    \caption{The 1-D distribution of 1.4~\msol{} star radius \ron{} from injected data set: {(1) NL3$\omega\rho$ and (2) TM1-2$\omega\rho$.} {The injected \ron value from {the chosen EOS} shown in black dashed line.} The red dashed lines represent the quantiles of each distribution at 16\%, 50\%, and 84\%. The density function is computed using KDE estimation.}
    \label{fig:injected_compare}
\end{figure*}
\subsection{Scenario 2}
Under the assumption that both J0030 and J0437 have a mass of $1.4 M_\odot$, we first apply Bayesian inference to determine the radius distribution for each sources. In Figure \ref{R1R2_posteiror}, we present the posterior distributions of $R_1$ and $R_2$, filtered by the measurements from J0030 and J0437. Here, we present the joint distribution of $R_1$ and $R_2$, which serve as the model parameters in this Bayesian inference. In our model, the two mass-radius points are fixed at $1.4 \, M_\odot$, with $R_1$ and $R_2$ as free parameters. These two parameters are individually fit to the likelihoods of the J0030 and J0437 observations, respectively, resulting in posterior distributions for each parameter.
We note that $R_1$ and $R_2$ are independent from each other (by definition) and therefore show no correlation in the 2D posterior. For J0437, the hotspot choice of J0030 cannot influence the $R_2$ posterior and this is also seen in Figure 2 (the small differences are due to sampling fluctuations between different runs). The results show that, since the same mass-radius measurement for J0437 is used across all models, the $R_2$ distributions for the three hotspot configurations overlap significantly. For J0030, however, the center of these radius posteriors shifts to larger values as we move from ST+PDT to ST-PST. The uncertainties increase from ST+PDT model to the PDT-U model, and then to the ST+PST model, reflecting the variation in posterior uncertainty among different models. Table \ref{Bayes_evidence} presents the Bayesian evidence for each configuration, indicating that among the models considered, the PDT-U model most closely represents a $1.4 M_\odot$ star, yielding a mass estimate of $M = 1.41^{+0.20}_{-0.19} M_\odot$, compared to the ST+PST model, which yields $1.34_{-0.16}^{+0.15} \, M_{\odot}$. Consistent with the observational data, the ST+PDT measurement is the least favored to represent a 1.4 \msol star in this NICER-only posterior, with the mass estimate being $M = 1.20_{-0.11}^{+0.14}$\msol.

If we assume both sources have a mass of $1.4 M_\odot$, we can calculate the joint posterior using Equation \ref{overall}. The $R_{1.4}$ distribution is shown in Figure \ref{Second_R14_compare}, which includes the probabilities that J0030 and J0437 are indeed $1.4 M_\odot$ stars. The plausibility of this inference is further reinforced, as deriving the distribution of \ron with mass measurements significantly different from $1.4 M_\odot$ could lead to misleading results. In Figure \ref{Second_R14_compare}, comparing the different probability distributions, we find that the PDT-U model yields a probability distribution for \ron that is higher than that derived from the ST+PDT mass-radius measurements, while the ST+PST model yields a distribution that is lower than that of ST+PDT. This pattern is consistent with the fact that the PDT-U model produces a mass-radius result for J0030 that is closest to $1.4 M_\odot$.

The values for $R_{1.4}$ reported in the ST+PDT case are $12.09_{-0.63}^{+0.81} \mathrm{~km}$, for ST+PST they are $12.67_{-0.86}^{+1.03} \mathrm{~km}$, and for PDT-U, $12.29_{-0.77}^{+1.02} \mathrm{~km}$. The uncertainties for these quantities are reflected in the posteriors shown in Figure \ref{R1R2_posteiror}. Note the uncertainties of \ron quoted here {are for the 95\% interval}.

All the results from the direct extraction of $R_{1.4}$ in both scenarios are summarized in Table \ref{R14_table}. Notably, all Scenario 2 results are highly consistent with each other within the 95\% credible range, preferring values larger than 12\,km. In contrast, for Scenario 1, all predictions are smaller than those in Scenario 2; however, the central values in Scenario 2 are still well within the uncertainty range of Scenario 1. Given the lower posterior uncertainty in radius applied in Scenario 2 with the ST+PDT hotspot model for J0030, this configuration produces the narrowest 95\% credible interval. Overall, since the PDT-U model predicts a mass for J0030 that is closest to 1.4 \msol, this setup yields the best prediction for \ron in this study, with the highest normalized Gaussian-like peak.

Notably, the estimations of \ron from Scenario 1 and Scenario 2 exhibit an inconsistent pattern, particularly with Scenario 1 providing a significantly larger estimation of \ron when using the PDT-U hotspot model. This can be understood by comparing the left panel of Figure 10 in \cite{Vinciguerra_2024}, which shows the NICER-only inference result for PDT-U, with the ST+PST and ST+PDT NICER-only results (see Figures 7 and 8 in the same paper). The mass-radius range for the PDT-U result is significantly broader and shifted toward the higher end. Consequently, when generating the pairs, more points fall into the larger mass-radius range, which generally leads to an increased prediction of \ron. Therefore, it is expected that the PDT-U result is larger than the ST+PDT and ST+PST results.

However, this trend is not observed in Scenario 2. This is because, in Scenario 2, the mass is fixed at $1.4 \, M_\odot$, meaning that the radius distribution of J0030, under a specific hotspot configuration and the $1.4 \, M_\odot$ condition, primarily determines the central value of \ron. Since the same J0437 result is used in both cases, this explains the absence of a similar trend in Scenario 2.

\section{Future case study: simulated dataset}
Currently, this study still suffers from systematic uncertainties in the inference. Compactness is the main factor influencing the X-ray pulse profile, so modeling this data results in a linearly correlated mass-radius measurement. This correlation is intrinsic to the modeling method of the observation itself. To overcome this, one possible approach is to introduce a strong mass prior, which would decouple this dependence, as in the case of J0437 \cite{Choudhury_2024}.

To showcase the power of our modeling method and to quantify to what extent our method can reproduce the underlying \ron, we extend our study to focus on simulated data. A similar simulation procedure has been discussed in \cite{Huang:2023grj}. We simulate two neutron star sources near the 1.4 \msol mass with the precision expected from future X-ray telescopes like Athena, STROBE-X, and eXTP \cite{2013sf2a.conf..447B,Watts_2016,extp_watts,strobex,froning2024strobe}. 

To simulate the mass-radius observations, we inject an equation of state model to produce the mass-radius observation data set. For comparison purposes, we choose two relativistic mean field RMF-based EOSs as the injected EOS: NL3\(\omega\rho\) and TM1-2\(\omega\rho\) \cite{Scurto_2023,PhysRevC.87.055801}. This choice could vary with different EOS frameworks, but since RMF naturally embeds causality and has been extensively explored to fit current astrophysical observations and nuclear experiments, it is appropriate for our study.

Two sources are selected to represent future mass-radius observations of two stars near 1.4 \msol. In the NL3\(\omega\rho\) model, we select two mass-radius pairs: \( M_1 = 1.37\, M_\odot \), \( R_1 = 13.75\, \mathrm{km} \) and \( M_2 = 1.51\, M_\odot \), \( R_2 = 13.83\, \mathrm{km} \). For TM1-2\(\omega\rho\), the simulated mass-radius pairs are \( M_1 = 1.21\, M_\odot \), \( R_1 = 13.18\, \mathrm{km} \) and \( M_2 = 1.42\, M_\odot \), \( R_2 = 13.34\, \mathrm{km} \). All the mass-radius pairs are simulated with a {$\pm$5\% uncertainty band as their 68\% range}, which is a reasonable uncertainty level expected to be achievable within 5--10 yr of future X-ray telescope observations.

In both of the injected EOS models, \ron is well defined, providing a reference for comparison with our inference results, allowing us to assess how well our method can recover \ron.

The results are presented in Figure~\ref{fig:injected_compare}. Here, we implemented the Scenario 1 method to infer \ron from the simulated data set. The injected value is indicated by the black dashed line, which closely overlaps with the peak of our inference result. In both the NL3\(\omega\rho\) and TM1-2\(\omega\rho\) cases, the residuals between our posterior peak value and the injected value are approximately 0.02\,km, with a $\sim$ 4\% uncertainty level. This demonstrates that if we can achieve high-quality measurements of stars mass-radius around 1.4 \msol with a 5\% uncertainty level, the method described in this work is capable of recovering the underlying \ron with considerable accuracy.

Typically, traditional EOS inference methods predict \ron with additional systematic errors arising from different choices in the underlying EOS construction. However, since the method presented here is {largely} EOS-independent—except for the systematic errors introduced by the observations and the theoretical modeling process of the mass-radius posterior from pulse profile modeling—it has been shown, in limited cases, to be capable of recovering \ron from underlying injected mass-radius measurements.

\section{Discussion}
\subsection{Comparison with astrophysical constraints}
In recent studies, the radius of a $1.4 M_\odot$ neutron star is constrained through an EOS-dependent approach based on astrophysical observations. Typically, the EOS parameters are first constrained, and then the corresponding radius range for a $1.4 M_\odot$ star is reconstructed. In a recent study by \cite{Rutherford_2024}, the most up-to-date constraint on $R_{1.4}$ was reported by combining all existing astrophysical data, including X-ray timing observations from NICER (J0030, J0437, and the $2.1 M_\odot$ star PSR J0740+6620), gravitational wave constraints, and new $\chi$EFT calculations up to 1.5 times nuclear saturation density. This combined dataset produced a much narrower 95\% credible interval for $R_{1.4}$. However, it is important to note that these results remain EOS model-dependent. Under the same observational dataset, different EOS models in this study, such as speed of sound (CS) and polytrope (PP),  result in different central values and credible intervals for $R_{1.4}$. 

The key advantage of the present study is its ability to mostly avoid dependence on specific EOS models. By improving the sampling method used here, we substantially decreased the need to explore the many degrees of freedom in EOS modeling. A notable difference is that the central values for the radius predictions obtained from multi-messenger constraints are generally larger than those reported in this work. This difference can be attributed to the fact that our analysis only weights the observations of J0030 and J0437 in determining \ron, while the multi-messenger approach incorporates additional observational factors. The internal theoretical structure of the EOS, which limits the possible shapes of the mass-radius curve, also likely contributes to the tighter constraints on \ron. Here, since the inference result is mostly EOS-independent, the systematic error introduced by choosing different EOS models has already been naturally absorbed into the reported uncertainty. This partly explains why the uncertainty in our results is generally larger than reported by EOS-based inference methods.

In comparison with the posterior predictions of \ron from X-ray observations constrained by EOS models that include first-order phase transitions, as will be discussed in \cite{Huang}, the values of \ron reported are $12.32_{-0.99}^{+0.95}$ km and $11.92_{-0.63}^{+1.08}$ km in the cases that with and without including phase transitions. These constraints are based on the inclusion of one more observation from PSR J0740+6620, and use the ST+PDT model for J0030 only. The predictions within the 95\% credible interval for the model without a phase transition are consistent with the results from Scenario 1 and Scenario 2 using the ST+PDT model for J0030. When a first-order phase transition is included in the neutron star, it predicts a larger central value for $R_{1.4}$, primarily because the phase transition softens the EOS around the relevant density, allowing it to better match the observational data.

\subsection{Comparison with C-REX and P-REX Experiments and Gravitational wave detection}
The radius of a $1.4 M_\odot$ neutron star can be directly related to results from nuclear experiments, such as C-REX and P-REX. Notably, the P-REX experiment, as discussed in \cite{PREX:2021umo}, suggested a relatively large $R_{1.4}$ value, with estimates of $R_{1.4} \gtrsim 13.25 \mathrm{~km}$ \cite{Reed_2021}. However, other studies \cite{Yue:2021yfx,Essick2021,Reinhard21} have reported smaller values for $R_{1.4}$. As noted by \cite{Reinhard21,Reinhard:2022inh,Mondal:2022cva}, a tension remains between experimental results and theoretical predictions. C-REX predicts a smaller and more comparable radius for a $1.4 M_\odot$ neutron star comparing with the result in this current study, highlighting the ongoing debate between different experimental approaches. This further underscores the importance of independently extracting $R_{1.4}$ from astrophysics side mostly excluding the dependence on any specific EOS model, as done in this study, to provide a useful point of comparison.

We also computed the tidal deformability ($\Lambda_{1.4}$) corresponding to the $R_{1.4}$ values. Using the empirical formula provided by \cite{Eemeli}, 
$
\Lambda(R_{1.4}) = 2.88 \times 10^{-6} \left( R_{1.4}/\mathrm{km} \right)^{7.5},
$
we estimated $\Lambda_{1.4}$ values and listed them in Table \ref{R14_table}. The GW170817 event reported $\Lambda_{1.4}=190_{-120}^{+390}$. Our computed values for $\Lambda_{1.4}$ are largely consistent with this range, suggesting that our results align with gravitational wave data. Furthermore, our findings, particularly for J0437, point towards a smaller radius and a softer EOS, consistent with results from \cite{Huang} and \cite{Rutherford_2024}.

\section{Conclusion}
In conclusion, this study presents a data-driven approach to determining the radius of a 1.4 \msol neutron star (\ron) by leveraging new NICER observations of PSR J0030+0451 and PSR J0437-4715. Unlike traditional methods that rely on a predefined equation of state (EOS), we demonstrate that \ron can be inferred directly from observational data, {mostly} independent of specific EOS models. This approach not only offers a novel way to constrain canonical neutron star properties but also provides insights into the tension surrounding neutron skin thickness measurements from nuclear experiments like C-REX and P-REX, by comparing the \ron predictions from these collaborations, even though the results could be model-dependent. By conducting the future case study presented in this work, we show that, in selected cases, the method we report can provide a reasonable prediction of \ron that closely matches the underlying injected \ron.

Our analysis highlights that different hotspot configurations for J0030 influence the mass-radius measurements and the inferred \ron. However, by comparing these configurations, we obtain consistent estimates of \ron across multiple scenarios. Furthermore, our results remain in agreement with constraints from gravitational wave data (e.g., GW170817), while providing a more flexible framework free from EOS assumptions. This work opens a new avenue for future observations, offering a robust method to extract neutron star properties directly from astrophysical data.
\section{Software and third party data repository citations} \label{sec:cite}

\textit{CompactObject package: An full-scope open-source Bayesian inference framework especially designed for Neutron star physics} Zenodo: DOI: 10.5281/zenodo.10927600 for Version 2.0.0, Github: \url{https://github.com/ChunHuangPhy/CompactObject}, documentation: \url{https://chunhuangphy.github.io/CompactObject/}. \textit{UltraNest}: \cite{2021JOSS....6.3001B}, \url{https://github.com/JohannesBuchner/UltraNest}.

\section{Acknowledgments}
C.H. acknowledges support from NASA grant 80NSSC24K1095. The author extends gratitude to Mark Alford, Alexander Chen, Ryan O'Connor and Chuyi Deng for insightful discussion. Thank the anonymous referee for their valuable feedback, which greatly contributed to improving the quality of this work.




\bibliography{main}{}

\begin{thebibliography}{}
\expandafter\ifx\csname natexlab\endcsname\relax\def\natexlab#1{#1}\fi
\providecommand{\url}[1]{\href{#1}{#1}}
\providecommand{\dodoi}[1]{doi:~\href{http://doi.org/#1}{\nolinkurl{#1}}}
\providecommand{\doeprint}[1]{\href{http://ascl.net/#1}{\nolinkurl{http://ascl.net/#1}}}
\providecommand{\doarXiv}[1]{\href{https://arxiv.org/abs/#1}{\nolinkurl{https://arxiv.org/abs/#1}}}

\bibitem[{Abbott {et~al.}(2018)Abbott, Abbott, Abbott, Acernese, Ackley, Adams, Adams, Addesso, Adhikari, Adya, Affeldt, Agarwal, Agathos, Agatsuma, Aggarwal, Aguiar, Aiello, Ain, Ajith, Allen, Allen, Allocca, Aloy, Altin, Amato, Ananyeva, Anderson, Anderson, Angelova, Antier, Appert, Arai, Araya, Areeda, Arène, Arnaud, Arun, Ascenzi, Ashton, Ast, Aston, Astone, Atallah, Aubin, Aufmuth, Aulbert, AultONeal, Austin, Avila-Alvarez, Babak, Bacon, Badaracco, Bader, Bae, Baker, Baldaccini, Ballardin, Ballmer, Banagiri, Barayoga, Barclay, Barish, Barker, Barkett, Barnum, Barone, Barr, Barsotti, Barsuglia, Barta, Bartlett, Bartos, Bassiri, Basti, Batch, Bawaj, Bayley, Bazzan, Bécsy, Beer, Bejger, Belahcene, Bell, Beniwal, Bensch, Berger, Bergmann, Bernuzzi, Bero, Berry, Bersanetti, Bertolini, Betzwieser, Bhandare, Bilenko, Bilgili, Billingsley, Billman, Birch, Birney, Birnholtz, Biscans, Biscoveanu, Bisht, Bitossi, Bizouard, Blackburn, Blackman, Blair, Blair, Blair, Bloemen, Bock, Bode, Boer, Boetzel, Bogaert,
  Bohe, Bondu, Bonilla, Bonnand, Booker, Boom, Booth, Bork, Boschi, Bose, Bossie, Bossilkov, Bosveld, Bouffanais, Bozzi, Bradaschia, Brady, Bramley, Branchesi, Brau, Briant, Brighenti, Brillet, Brinkmann, Brisson, Brockill, Brooks, Brown, Brunett, Buchanan, Buikema, Bulik, Bulten, Buonanno, Buskulic, Buy, Byer, Cabero, Cadonati, Cagnoli, Cahillane, Calderón~Bustillo, Callister, Calloni, Camp, Canepa, Canizares, Cannon, Cao, Cao, Capano, Capocasa, Carbognani, Caride, Carney, Carullo, Casanueva~Diaz, Casentini, Caudill, Cavaglià, Cavalier, Cavalieri, Cella, Cepeda, Cerdá-Durán, Cerretani, Cesarini, Chaibi, Chamberlin, Chan, Chao, Charlton, Chase, Chassande-Mottin, Chatterjee, Chatziioannou, Cheeseboro, Chen, Chen, Chen, Cheng, Chia, Chincarini, Chiummo, Chmiel, Cho, Cho, Chow, Christensen, Chu, Chua, Chua, Chung, Chung, Ciani, Ciobanu, Ciolfi, Cipriano, Cirelli, Cirone, Clara, Clark, Clearwater, Cleva, Cocchieri, Coccia, Cohadon, Cohen, Colla, Collette, Collins, Cominsky, Constancio, Conti, Cooper, Corban,
  Corbitt, Cordero-Carrión, Corley, Cornish, Corsi, Cortese, Costa, Cotesta, Coughlin, Coughlin, Coulon, Countryman, Couvares, Covas, Cowan, Coward, Cowart, Coyne, Coyne, Creighton, Creighton, Cripe, Crowder, Cullen, Cumming, Cunningham, Cuoco, Canton, Dálya, Danilishin, D’Antonio, Danzmann, Dasgupta, Da~Silva~Costa, Dattilo, Dave, Davier, Davis, Daw, Day, DeBra, Deenadayalan, Degallaix, De~Laurentis, Deléglise, Del~Pozzo, Demos, Denker, Dent, De~Pietri, Derby, Dergachev, De~Rosa, De~Rossi, DeSalvo, de~Varona, Dhurandhar, Díaz, Dietrich, Di~Fiore, Di~Giovanni, Di~Girolamo, Di~Lieto, Ding, Di~Pace, Di~Palma, Di~Renzo, Dmitriev, Doctor, Dolique, Donovan, Dooley, Doravari, Dorrington, Dovale~Álvarez, Downes, Drago, Dreissigacker, Driggers, Du, Dupej, Dwyer, Easter, Edo, Edwards, Effler, Eggenstein, Ehrens, Eichholz, Eikenberry, Eisenmann, Eisenstein, Essick, Estelles, Estevez, Etienne, Etzel, Evans, Evans, Fafone, Fair, Fairhurst, Fan, Farinon, Farr, Farr, Fauchon-Jones, Favata, Fays, Fee, Fehrmann,
  Feicht, Fejer, Feng, Fernandez-Galiana, Ferrante, Ferreira, Ferrini, Fidecaro, Fiori, Fiorucci, Fishbach, Fisher, Fishner, Fitz-Axen, Flaminio, Fletcher, Fong, Font, Forsyth, Forsyth, Fournier, Frasca, Frasconi, Frei, Freise, Frey, Frey, Fritschel, Frolov, Fulda, Fyffe, Gabbard, Gadre, Gaebel, Gair, Gammaitoni, Ganija, Gaonkar, Garcia, García-Quirós, Garufi, Gateley, Gaudio, Gaur, Gayathri, Gemme, Genin, Gennai, George, George, Gergely, Germain, Ghonge, Ghosh, Ghosh, Ghosh, Giacomazzo, Giaime, Giardina, Giazotto, Gill, Giordano, Glover, Goetz, Goetz, Goncharov, González, Gonzalez~Castro, Gopakumar, Gorodetsky, Gossan, Gosselin, Gouaty, Grado, Graef, Granata, Grant, Gras, Gray, Greco, Green, Green, Gretarsson, Groot, Grote, Grunewald, Gruning, Guidi, Gulati, Guo, Gupta, Gupta, Gushwa, Gustafson, Gustafson, Halim, Hall, Hall, Hamilton, Hamilton, Hammond, Haney, Hanke, Hanks, Hanna, Hannam, Hannuksela, Hanson, Hardwick, Harms, Harry, Harry, Hart, Haster, Haughian, Healy, Heidmann, Heintze, Heitmann, Hello,
  Hemming, Hendry, Heng, Hennig, Heptonstall, Hernandez, Heurs, Hild, Hinderer, Ho, Hoak, Hochheim, Hofman, Holland, Holt, Holz, Hopkins, Horst, Hough, Houston, Howell, Hreibi, Huerta, Huet, Hughey, Hulko, Husa, Huttner, Huynh-Dinh, Iess, Indik, Ingram, Inta, Intini, Irwin, Isa, Isac, Isi, Iyer, Izumi, Jacqmin, Jani, Jaranowski, Johnson, Johnson, Jones, Jones, Jonker, Ju, Junker, Kalaghatgi, Kalogera, Kamai, Kandhasamy, Kang, Kanner, Kapadia, Karki, Karvinen, Kasprzack, Katolik, Katsanevas, Katsavounidis, Katzman, Kaufer, Kawabe, Keerthana, Kéfélian, Keitel, Kemball, Kennedy, Key, Khalili, Khamesra, Khan, Khan, Khan, Khan, Khazanov, Kijbunchoo, Kim, Kim, Kim, Kim, Kim, Kim, King, King, Kinley-Hanlon, Kirchhoff, Kissel, Kleybolte, Klimenko, Knowles, Koch, Koehlenbeck, Koley, Kondrashov, Kontos, Korobko, Korth, Kowalska, Kozak, Krämer, Kringel, Krishnan, Królak, Kuehn, Kumar, Kumar, Kumar, Kuo, Kutynia, Kwang, Lackey, Lai, Landry, Landry, Lang, Lange, Lantz, Lanza, Lartaux-Vollard, Lasky, Laxen, Lazzarini,
  Lazzaro, Leaci, Leavey, Lee, Lee, Lee, Lee, Lee, Lehmann, Lenon, Leonardi, Leroy, Letendre, Levin, Li, Li, Li, Linker, Littenberg, Liu, Liu, Lo, Lockerbie, London, Longo, Lorenzini, Loriette, Lormand, Losurdo, Lough, Lousto, Lovelace, Lück, Lumaca, Lundgren, Lynch, Ma, Macas, Macfoy, Machenschalk, MacInnis, Macleod, Magaña~Hernandez, Magaña-Sandoval, Magaña~Zertuche, Magee, Majorana, Maksimovic, Man, Mandic, Mangano, Mansell, Manske, Mantovani, Marchesoni, Marion, Márka, Márka, Markakis, Markosyan, Markowitz, Maros, Marquina, Martelli, Martellini, Martin, Martin, Martynov, Mason, Massera, Masserot, Massinger, Masso-Reid, Mastrogiovanni, Matas, Matichard, Matone, Mavalvala, Mazumder, McCann, McCarthy, McClelland, McCormick, McCuller, McGuire, McIver, McManus, McRae, McWilliams, Meacher, Meadors, Mehmet, Meidam, Mejuto-Villa, Melatos, Mendell, Mendoza-Gandara, Mercer, Mereni, Merilh, Merzougui, Meshkov, Messenger, Messick, Metzdorff, Meyers, Miao, Michel, Middleton, Mikhailov, Milano, Miller, Miller,
  Miller, Miller, Millhouse, Mills, Milovich-Goff, Minazzoli, Minenkov, Ming, Mishra, Mitra, Mitrofanov, Mitselmakher, Mittleman, Moffa, Mogushi, Mohan, Mohapatra, Montani, Moore, Moraru, Moreno, Morisaki, Mours, Mow-Lowry, Mueller, Muir, Mukherjee, Mukherjee, Mukherjee, Mukund, Mullavey, Munch, Muñiz, Muratore, Murray, Nagar, Napier, Nardecchia, Naticchioni, Nayak, Neilson, Nelemans, Nelson, Nery, Neunzert, Nevin, Newport, Ng, Ng, Nguyen, Nguyen, Nichols, Nielsen, Nissanke, Nitz, Nocera, Nolting, North, Nuttall, Obergaulinger, Oberling, O’Brien, O’Dea, Ogin, Oh, Oh, Ohme, Ohta, Okada, Oliver, Oppermann, Oram, O’Reilly, Ormiston, Ortega, O’Shaughnessy, Ossokine, Ottaway, Overmier, Owen, Pace, Pagano, Page, Page, Pai, Pai, Palamos, Palashov, Palomba, Pal-Singh, Pan, Pan, Pang, Pang, Pankow, Pannarale, Pant, Paoletti, Paoli, Papa, Parida, Parker, Pascucci, Pasqualetti, Passaquieti, Passuello, Patil, Patricelli, Pearlstone, Pedersen, Pedraza, Pedurand, Pekowsky, Pele, Penn, Perego, Perez, Perreca,
  Perri, Pfeiffer, Phelps, Phukon, Piccinni, Pichot, Piergiovanni, Pierro, Pillant, Pinard, Pinto, Pirello, Pitkin, Poggiani, Popolizio, Porter, Possenti, Post, Powell, Prasad, Pratt, Pratten, Predoi, Prestegard, Principe, Privitera, Prodi, Prokhorov, Puncken, Punturo, Puppo, Pürrer, Qi, Quetschke, Quintero, Quitzow-James, Raab, Rabeling, Radkins, Raffai, Raja, Rajan, Rajbhandari, Rakhmanov, Ramirez, Ramos-Buades, Rana, Rapagnani, Raymond, Razzano, Read, Regimbau, Rei, Reid, Reitze, Ren, Ricci, Ricker, Riemenschneider, Riles, Rizzo, Robertson, Robie, Robinet, Robson, Rocchi, Rolland, Rollins, Roma, Romano, Romel, Romie, Rosińska, Ross, Rowan, Rüdiger, Ruggi, Rutins, Ryan, Sachdev, Sadecki, Sakellariadou, Salconi, Saleem, Salemi, Samajdar, Sammut, Sampson, Sanchez, Sanchez, Sanchis-Gual, Sandberg, Sanders, Sarin, Sassolas, Sathyaprakash, Saulson, Sauter, Savage, Sawadsky, Schale, Scheel, Scheuer, Schmidt, Schnabel, Schofield, Schönbeck, Schreiber, Schuette, Schulte, Schutz, Schwalbe, Scott, Scott, Seidel,
  Sellers, Sengupta, Sentenac, Sequino, Sergeev, Setyawati, Shaddock, Shaffer, Shah, Shahriar, Shaner, Shao, Shapiro, Shawhan, Shen, Shoemaker, Shoemaker, Siellez, Siemens, Sieniawska, Sigg, Silva, Singer, Singh, Singhal, Sintes, Slagmolen, Slaven-Blair, Smith, Smith, Smith, Somala, Son, Sorazu, Sorrentino, Souradeep, Spencer, Srivastava, Staats, Steinke, Steinlechner, Steinlechner, Steinmeyer, Steltner, Stevenson, Stocks, Stone, Stops, Strain, Stratta, Strigin, Strunk, Sturani, Stuver, Summerscales, Sun, Sunil, Suresh, Sutton, Swinkels, Szczepańczyk, Tacca, Tait, Talbot, Talukder, Tanner, Tápai, Taracchini, Tasson, Taylor, Taylor, Tewari, Theeg, Thies, Thomas, Thomas, Thomas, Thorne, Thrane, Tiwari, Tiwari, Tokmakov, Toland, Tonelli, Tornasi, Torres-Forné, Torrie, Töyrä, Travasso, Traylor, Trinastic, Tringali, Trovato, Trozzo, Tsang, Tse, Tso, Tsuna, Tsukada, Tuyenbayev, Ueno, Ugolini, Urban, Usman, Vahlbruch, Vajente, Valdes, van Bakel, van Beuzekom, van~den Brand, Van Den~Broeck, Vander-Hyde, van~der
  Schaaf, van Heijningen, van Veggel, Vardaro, Varma, Vass, Vasúth, Vecchio, Vedovato, Veitch, Veitch, Venkateswara, Venugopalan, Verkindt, Vetrano, Viceré, Viets, Vinciguerra, Vine, Vinet, Vitale, Vo, Vocca, Vorvick, Vyatchanin, Wade, Wade, Wade, Walet, Walker, Wallace, Walsh, Wang, Wang, Wang, Wang, Wang, Ward, Warner, Was, Watchi, Weaver, Wei, Weinert, Weinstein, Weiss, Wellmann, Wen, Wessel, Weßels, Westerweck, Wette, Whelan, Whiting, Whittle, Wilken, Williams, Williams, Williamson, Willis, Willke, Wimmer, Winkler, Wipf, Wittel, Woan, Woehler, Wofford, Wong, Worden, Wright, Wu, Wysocki, Xiao, Yam, Yamamoto, Yancey, Yang, Yap, Yazback, Yu, Yu, Yvert, Zadrożny, Zanolin, Zelenova, Zendri, Zevin, Zhang, Zhang, Zhang, Zhang, Zhang, Zhao, Zhou, Zhou, Zhu, Zhu, Zimmerman, Zlochower, Zucker, \& Zweizig}]{Abbott_2018}
Abbott, B., Abbott, R., Abbott, T., {et~al.} 2018, Physical Review Letters, 121, \dodoi{10.1103/physrevlett.121.161101}

\bibitem[{Abbott {et~al.}(2017)Abbott, Abbott, Abbott, Acernese, Ackley, Adams, Adams, Addesso, Adhikari, Adya, Affeldt, Afrough, Agarwal, Agathos, Agatsuma, Aggarwal, Aguiar, Aiello, Ain, Ajith, Allen, Allen, Allocca, Altin, Amato, Ananyeva, Anderson, Anderson, Angelova, Antier, Appert, Arai, Araya, Areeda, Arnaud, Arun, Ascenzi, Ashton, Ast, Aston, Astone, Atallah, Aufmuth, Aulbert, AultONeal, Austin, Avila-Alvarez, Babak, Bacon, Bader, Bae, Bailes, Baker, Baldaccini, Ballardin, Ballmer, Banagiri, Barayoga, Barclay, Barish, Barker, Barkett, Barone, Barr, Barsotti, Barsuglia, Barta, Barthelmy, Bartlett, Bartos, Bassiri, Basti, Batch, Bawaj, Bayley, Bazzan, Bécsy, Beer, Bejger, Belahcene, Bell, Berger, Bergmann, Bernuzzi, Bero, Berry, Bersanetti, Bertolini, Betzwieser, Bhagwat, Bhandare, Bilenko, Billingsley, Billman, Birch, Birney, Birnholtz, Biscans, Biscoveanu, Bisht, Bitossi, Biwer, Bizouard, Blackburn, Blackman, Blair, Blair, Blair, Bloemen, Bock, Bode, Boer, Bogaert, Bohe, Bondu, Bonilla, Bonnand,
  Boom, Bork, Boschi, Bose, Bossie, Bouffanais, Bozzi, Bradaschia, Brady, Branchesi, Brau, Briant, Brillet, Brinkmann, Brisson, Brockill, Broida, Brooks, Brown, Brown, Brunett, Buchanan, Buikema, Bulik, Bulten, Buonanno, Buskulic, Buy, Byer, Cabero, Cadonati, Cagnoli, Cahillane, Calderón~Bustillo, Callister, Calloni, Camp, Canepa, Canizares, Cannon, Cao, Cao, Capano, Capocasa, Carbognani, Caride, Carney, Carullo, Casanueva~Diaz, Casentini, Caudill, Cavaglià, Cavalier, Cavalieri, Cella, Cepeda, Cerdá-Durán, Cerretani, Cesarini, Chamberlin, Chan, Chao, Charlton, Chase, Chassande-Mottin, Chatterjee, Chatziioannou, Cheeseboro, Chen, Chen, Chen, Cheng, Chia, Chincarini, Chiummo, Chmiel, Cho, Cho, Chow, Christensen, Chu, Chua, Chua, Chung, Chung, Ciani, Ciolfi, Cirelli, Cirone, Clara, Clark, Clearwater, Cleva, Cocchieri, Coccia, Cohadon, Cohen, Colla, Collette, Cominsky, Constancio, Conti, Cooper, Corban, Corbitt, Cordero-Carrión, Corley, Cornish, Corsi, Cortese, Costa, Coughlin, Coughlin, Coulon, Countryman,
  Couvares, Covas, Cowan, Coward, Cowart, Coyne, Coyne, Creighton, Creighton, Cripe, Crowder, Cullen, Cumming, Cunningham, Cuoco, Dal~Canton, Dálya, Danilishin, D’Antonio, Danzmann, Dasgupta, Da~Silva~Costa, Dattilo, Dave, Davier, Davis, Daw, Day, De, DeBra, Degallaix, De~Laurentis, Deléglise, Del~Pozzo, Demos, Denker, Dent, De~Pietri, Dergachev, De~Rosa, DeRosa, De~Rossi, DeSalvo, de~Varona, Devenson, Dhurandhar, Díaz, Dietrich, Di~Fiore, Di~Giovanni, Di~Girolamo, Di~Lieto, Di~Pace, Di~Palma, Di~Renzo, Doctor, Dolique, Donovan, Dooley, Doravari, Dorrington, Douglas, Dovale~Álvarez, Downes, Drago, Dreissigacker, Driggers, Du, Ducrot, Dudi, Dupej, Dwyer, Edo, Edwards, Effler, Eggenstein, Ehrens, Eichholz, Eikenberry, Eisenstein, Essick, Estevez, Etienne, Etzel, Evans, Evans, Factourovich, Fafone, Fair, Fairhurst, Fan, Farinon, Farr, Farr, Fauchon-Jones, Favata, Fays, Fee, Fehrmann, Feicht, Fejer, Fernandez-Galiana, Ferrante, Ferreira, Ferrini, Fidecaro, Finstad, Fiori, Fiorucci, Fishbach, Fisher,
  Fitz-Axen, Flaminio, Fletcher, Fong, Font, Forsyth, Forsyth, Fournier, Frasca, Frasconi, Frei, Freise, Frey, Frey, Fries, Fritschel, Frolov, Fulda, Fyffe, Gabbard, Gadre, Gaebel, Gair, Gammaitoni, Ganija, Gaonkar, Garcia-Quiros, Garufi, Gateley, Gaudio, Gaur, Gayathri, Gehrels, Gemme, Genin, Gennai, George, George, Gergely, Germain, Ghonge, Ghosh, Ghosh, Ghosh, Giaime, Giardina, Giazotto, Gill, Glover, Goetz, Goetz, Gomes, Goncharov, González, Gonzalez~Castro, Gopakumar, Gorodetsky, Gossan, Gosselin, Gouaty, Grado, Graef, Granata, Grant, Gras, Gray, Greco, Green, Gretarsson, Groot, Grote, Grunewald, Gruning, Guidi, Guo, Gupta, Gupta, Gushwa, Gustafson, Gustafson, Halim, Hall, Hall, Hamilton, Hammond, Haney, Hanke, Hanks, Hanna, Hannam, Hannuksela, Hanson, Hardwick, Harms, Harry, Harry, Hart, Haster, Haughian, Healy, Heidmann, Heintze, Heitmann, Hello, Hemming, Hendry, Heng, Hennig, Heptonstall, Heurs, Hild, Hinderer, Ho, Hoak, Hofman, Holt, Holz, Hopkins, Horst, Hough, Houston, Howell, Hreibi, Hu, Huerta,
  Huet, Hughey, Husa, Huttner, Huynh-Dinh, Indik, Inta, Intini, Isa, Isac, Isi, Iyer, Izumi, Jacqmin, Jani, Jaranowski, Jawahar, Jiménez-Forteza, Johnson, Johnson-McDaniel, Jones, Jones, Jonker, Ju, Junker, Kalaghatgi, Kalogera, Kamai, Kandhasamy, Kang, Kanner, Kapadia, Karki, Karvinen, Kasprzack, Kastaun, Katolik, Katsavounidis, Katzman, Kaufer, Kawabe, Kéfélian, Keitel, Kemball, Kennedy, Kent, Key, Khalili, Khan, Khan, Khan, Khazanov, Kijbunchoo, Kim, Kim, Kim, Kim, Kim, Kim, Kimbrell, King, King, Kinley-Hanlon, Kirchhoff, Kissel, Kleybolte, Klimenko, Knowles, Koch, Koehlenbeck, Koley, Kondrashov, Kontos, Korobko, Korth, Kowalska, Kozak, Krämer, Kringel, Krishnan, Królak, Kuehn, Kumar, Kumar, Kumar, Kuo, Kutynia, Kwang, Lackey, Lai, Landry, Lang, Lange, Lantz, Lanza, Larson, Lartaux-Vollard, Lasky, Laxen, Lazzarini, Lazzaro, Leaci, Leavey, Lee, Lee, Lee, Lee, Lee, Lehmann, Lenon, Leon, Leonardi, Leroy, Letendre, Levin, Li, Linker, Littenberg, Liu, Liu, Lo, Lockerbie, London, Lord, Lorenzini, Loriette,
  Lormand, Losurdo, Lough, Lousto, Lovelace, Lück, Lumaca, Lundgren, Lynch, Ma, Macas, Macfoy, Machenschalk, MacInnis, Macleod, Magaña~Hernandez, Magaña-Sandoval, Magaña~Zertuche, Magee, Majorana, Maksimovic, Man, Mandic, Mangano, Mansell, Manske, Mantovani, Marchesoni, Marion, Márka, Márka, Markakis, Markosyan, Markowitz, Maros, Marquina, Marsh, Martelli, Martellini, Martin, Martin, Martynov, Marx, Mason, Massera, Masserot, Massinger, Masso-Reid, Mastrogiovanni, Matas, Matichard, Matone, Mavalvala, Mazumder, McCarthy, McClelland, McCormick, McCuller, McGuire, McIntyre, McIver, McManus, McNeill, McRae, McWilliams, Meacher, Meadors, Mehmet, Meidam, Mejuto-Villa, Melatos, Mendell, Mercer, Merilh, Merzougui, Meshkov, Messenger, Messick, Metzdorff, Meyers, Miao, Michel, Middleton, Mikhailov, Milano, Miller, Miller, Miller, Millhouse, Milovich-Goff, Minazzoli, Minenkov, Ming, Mishra, Mitra, Mitrofanov, Mitselmakher, Mittleman, Moffa, Moggi, Mogushi, Mohan, Mohapatra, Molina, Montani, Moore, Moraru, Moreno,
  Morisaki, Morriss, Mours, Mow-Lowry, Mueller, Muir, Mukherjee, Mukherjee, Mukherjee, Mukund, Mullavey, Munch, Muñiz, Muratore, Murray, Nagar, Napier, Nardecchia, Naticchioni, Nayak, Neilson, Nelemans, Nelson, Nery, Neunzert, Nevin, Newport, Newton, Ng, Nguyen, Nguyen, Nichols, Nielsen, Nissanke, Nitz, Noack, Nocera, Nolting, North, Nuttall, Oberling, O’Dea, Ogin, Oh, Oh, Ohme, Okada, Oliver, Oppermann, Oram, O’Reilly, Ormiston, Ortega, O’Shaughnessy, Ossokine, Ottaway, Overmier, Owen, Pace, Page, Page, Pai, Pai, Palamos, Palashov, Palomba, Pal-Singh, Pan, Pan, Pang, Pang, Pankow, Pannarale, Pant, Paoletti, Paoli, Papa, Parida, Parker, Pascucci, Pasqualetti, Passaquieti, Passuello, Patil, Patricelli, Pearlstone, Pedraza, Pedurand, Pekowsky, Pele, Penn, Perez, Perreca, Perri, Pfeiffer, Phelps, Piccinni, Pichot, Piergiovanni, Pierro, Pillant, Pinard, Pinto, Pirello, Pitkin, Poe, Poggiani, Popolizio, Porter, Post, Powell, Prasad, Pratt, Pratten, Predoi, Prestegard, Prijatelj, Principe, Privitera, Prix,
  Prodi, Prokhorov, Puncken, Punturo, Puppo, Pürrer, Qi, Quetschke, Quintero, Quitzow-James, Raab, Rabeling, Radkins, Raffai, Raja, Rajan, Rajbhandari, Rakhmanov, Ramirez, Ramos-Buades, Rapagnani, Raymond, Razzano, Read, Regimbau, Rei, Reid, Reitze, Ren, Reyes, Ricci, Ricker, Rieger, Riles, Rizzo, Robertson, Robie, Robinet, Rocchi, Rolland, Rollins, Roma, Romano, Romano, Romel, Romie, Rosińska, Ross, Rowan, Rüdiger, Ruggi, Rutins, Ryan, Sachdev, Sadecki, Sadeghian, Sakellariadou, Salconi, Saleem, Salemi, Samajdar, Sammut, Sampson, Sanchez, Sanchez, Sanchis-Gual, Sandberg, Sanders, Sassolas, Sathyaprakash, Saulson, Sauter, Savage, Sawadsky, Schale, Scheel, Scheuer, Schmidt, Schmidt, Schnabel, Schofield, Schönbeck, Schreiber, Schuette, Schulte, Schutz, Schwalbe, Scott, Scott, Seidel, Sellers, Sengupta, Sentenac, Sequino, Sergeev, Shaddock, Shaffer, Shah, Shahriar, Shaner, Shao, Shapiro, Shawhan, Sheperd, Shoemaker, Shoemaker, Siellez, Siemens, Sieniawska, Sigg, Silva, Singer, Singh, Singhal, Sintes,
  Slagmolen, Smith, Smith, Smith, Somala, Son, Sonnenberg, Sorazu, Sorrentino, Souradeep, Spencer, Srivastava, Staats, Staley, Steinke, Steinlechner, Steinlechner, Steinmeyer, Stevenson, Stone, Stops, Strain, Stratta, Strigin, Strunk, Sturani, Stuver, Summerscales, Sun, Sunil, Suresh, Sutton, Swinkels, Szczepańczyk, Tacca, Tait, Talbot, Talukder, Tanner, Tápai, Taracchini, Tasson, Taylor, Taylor, Tewari, Theeg, Thies, Thomas, Thomas, Thomas, Thorne, Thorne, Thrane, Tiwari, Tiwari, Tokmakov, Toland, Tonelli, Tornasi, Torres-Forné, Torrie, Töyrä, Travasso, Traylor, Trinastic, Tringali, Trozzo, Tsang, Tse, Tso, Tsukada, Tsuna, Tuyenbayev, Ueno, Ugolini, Unnikrishnan, Urban, Usman, Vahlbruch, Vajente, Valdes, Vallisneri, van Bakel, van Beuzekom, van~den Brand, Van Den~Broeck, Vander-Hyde, van~der Schaaf, van Heijningen, van Veggel, Vardaro, Varma, Vass, Vasúth, Vecchio, Vedovato, Veitch, Veitch, Venkateswara, Venugopalan, Verkindt, Vetrano, Viceré, Viets, Vinciguerra, Vine, Vinet, Vitale, Vo, Vocca,
  Vorvick, Vyatchanin, Wade, Wade, Wade, Walet, Walker, Wallace, Walsh, Wang, Wang, Wang, Wang, Wang, Ward, Warner, Was, Watchi, Weaver, Wei, Weinert, Weinstein, Weiss, Wen, Wessel, Weßels, Westerweck, Westphal, Wette, Whelan, Whitcomb, Whiting, Whittle, Wilken, Williams, Williams, Williamson, Willis, Willke, Wimmer, Winkler, Wipf, Wittel, Woan, Woehler, Wofford, Wong, Worden, Wright, Wu, Wysocki, Xiao, Yamamoto, Yancey, Yang, Yap, Yazback, Yu, Yu, Yvert, Zadrożny, Zanolin, Zelenova, Zendri, Zevin, Zhang, Zhang, Zhang, Zhang, Zhao, Zhou, Zhou, Zhu, Zhu, Zimmerman, Zucker, \& Zweizig}]{Abbott_2017}
Abbott, B.~P., Abbott, R., Abbott, T.~D., {et~al.} 2017, Physical Review Letters, 119, \dodoi{10.1103/physrevlett.119.161101}

\bibitem[{Abbott {et~al.}(2019)Abbott, Abbott, Abbott, Acernese, Ackley, Adams, Adams, Addesso, Adhikari, Adya, Affeldt, Agarwal, Agathos, Agatsuma, Aggarwal, Aguiar, Aiello, Ain, Ajith, Allen, Allen, Allocca, Aloy, Altin, Amato, Ananyeva, Anderson, Anderson, Angelova, Antier, Appert, Arai, Araya, Areeda, Ar\`ene, Arnaud, Arun, Ascenzi, Ashton, Ast, Aston, Astone, Atallah, Aubin, Aufmuth, Aulbert, AultONeal, Austin, Avila-Alvarez, Babak, Bacon, Badaracco, Bader, Bae, Baker, Baldaccini, Ballardin, Ballmer, Banagiri, Barayoga, Barclay, Barish, Barker, Barkett, Barnum, Barone, Barr, Barsotti, Barsuglia, Barta, Bartlett, Bartos, Bassiri, Basti, Batch, Bawaj, Bayley, Bazzan, B\'ecsy, Beer, Bejger, Belahcene, Bell, Beniwal, Bensch, Berger, Bergmann, Bernuzzi, Bero, Berry, Bersanetti, Bertolini, Betzwieser, Bhandare, Bilenko, Bilgili, Billingsley, Billman, Birch, Birney, Birnholtz, Biscans, Biscoveanu, Bisht, Bitossi, Bizouard, Blackburn, Blackman, Blair, Blair, Blair, Bloemen, Bock, Bode, Boer, Boetzel, Bogaert,
  Bohe, Bondu, Bonilla, Bonnand, Booker, Boom, Booth, Bork, Boschi, Bose, Bossie, Bossilkov, Bosveld, Bouffanais, Bozzi, Bradaschia, Brady, Bramley, Branchesi, Brau, Briant, Brighenti, Brillet, Brinkmann, Brisson, Brockill, Brooks, Brown, Brunett, Buchanan, Buikema, Bulik, Bulten, Buonanno, Buskulic, Buy, Byer, Cabero, Cadonati, Cagnoli, Cahillane, Bustillo, Callister, Calloni, Camp, Canepa, Canizares, Cannon, Cao, Cao, Capano, Capocasa, Carbognani, Caride, Carney, Carullo, Diaz, Casentini, Caudill, Cavagli\`a, Cavalier, Cavalieri, Cella, Cepeda, Cerd\'a-Dur\'an, Cerretani, Cesarini, Chaibi, Chamberlin, Chan, Chao, Charlton, Chase, Chassande-Mottin, Chatterjee, Chatziioannou, Cheeseboro, Chen, Chen, Chen, Cheng, Chia, Chincarini, Chiummo, Chmiel, Cho, Cho, Chow, Christensen, Chu, Chua, Chua, Chung, Chung, Ciani, Ciobanu, Ciolfi, Cipriano, Cirelli, Cirone, Clara, Clark, Clearwater, Cleva, Cocchieri, Coccia, Cohadon, Cohen, Colla, Collette, Collins, Cominsky, Constancio, Conti, Cooper, Corban, Corbitt,
  Cordero-Carri\'on, Corley, Cornish, Corsi, Cortese, Costa, Cotesta, Coughlin, Coughlin, Coulon, Countryman, Couvares, Covas, Cowan, Coward, Cowart, Coyne, Coyne, Creighton, Creighton, Cripe, Crowder, Cullen, Cumming, Cunningham, Cuoco, Canton, D\'alya, Danilishin, D'Antonio, Danzmann, Dasgupta, Costa, Dattilo, Dave, Davier, Davis, Daw, Day, DeBra, Deenadayalan, Degallaix, De~Laurentis, Del\'eglise, Del~Pozzo, Demos, Denker, Dent, De~Pietri, Derby, Dergachev, De~Rosa, De~Rossi, DeSalvo, de~Varona, Dhurandhar, D\'{\i}az, Dietrich, Di~Fiore, Di~Giovanni, Di~Girolamo, Di~Lieto, Ding, Di~Pace, Di~Palma, Di~Renzo, Dmitriev, Doctor, Dolique, Donovan, Dooley, Doravari, Dorrington, \'Alvarez, Downes, Drago, Dreissigacker, Driggers, Du, Dudi, Dupej, Dwyer, Easter, Edo, Edwards, Effler, Eggenstein, Ehrens, Eichholz, Eikenberry, Eisenmann, Eisenstein, Essick, Estelles, Estevez, Etienne, Etzel, Evans, Evans, Fafone, Fair, Fairhurst, Fan, Farinon, Farr, Farr, Fauchon-Jones, Favata, Fays, Fee, Fehrmann, Feicht, Fejer,
  Feng, Fernandez-Galiana, Ferrante, Ferreira, Ferrini, Fidecaro, Fiori, Fiorucci, Fishbach, Fisher, Fishner, Fitz-Axen, Flaminio, Fletcher, Fong, Font, Forsyth, Forsyth, Fournier, Frasca, Frasconi, Frei, Freise, Frey, Frey, Fritschel, Frolov, Fulda, Fyffe, Gabbard, Gadre, Gaebel, Gair, Gammaitoni, Ganija, Gaonkar, Garcia, Garc\'{\i}a-Quir\'os, Garufi, Gateley, Gaudio, Gaur, Gayathri, Gemme, Genin, Gennai, George, George, Gergely, Germain, Ghonge, Ghosh, Ghosh, Ghosh, Giacomazzo, Giaime, Giardina, Giazotto, Gill, Giordano, Glover, Goetz, Goetz, Goncharov, Gonz\'alez, Castro, Gopakumar, Gorodetsky, Gossan, Gosselin, Gouaty, Grado, Graef, Granata, Grant, Gras, Gray, Greco, Green, Green, Gretarsson, Groot, Grote, Grunewald, Gruning, Guidi, Gulati, Guo, Gupta, Gupta, Gushwa, Gustafson, Gustafson, Halim, Hall, Hall, Hamilton, Hamilton, Hammond, Haney, Hanke, Hanks, Hanna, Hannam, Hannuksela, Hanson, Hardwick, Harms, Harry, Harry, Hart, Haster, Haughian, Healy, Heidmann, Heintze, Heitmann, Hello, Hemming, Hendry,
  Heng, Hennig, Heptonstall, Hernandez, Heurs, Hild, Hinderer, Hoak, Hochheim, Hofman, Holland, Holt, Holz, Hopkins, Horst, Hough, Houston, Howell, Hreibi, Huerta, Huet, Hughey, Hulko, Husa, Huttner, Huynh-Dinh, Iess, Indik, Ingram, Inta, Intini, Isa, Isac, Isi, Iyer, Izumi, Jacqmin, Jani, Jaranowski, Johnson, Johnson, Jones, Jones, Jonker, Ju, Junker, Kalaghatgi, Kalogera, Kamai, Kandhasamy, Kang, Kanner, Kapadia, Karki, Karvinen, Kasprzack, Kastaun, Katolik, Katsanevas, Katsavounidis, Katzman, Kaufer, Kawabe, Keerthana, K\'ef\'elian, Keitel, Kemball, Kennedy, Key, Khalili, Khamesra, Khan, Khan, Khan, Khan, Khazanov, Kijbunchoo, Kim, Kim, Kim, Kim, Kim, Kim, King, King, Kinley-Hanlon, Kirchhoff, Kissel, Kleybolte, Klimenko, Knowles, Koch, Koehlenbeck, Koley, Kondrashov, Kontos, Korobko, Korth, Kowalska, Kozak, Kr\"amer, Kringel, Krishnan, Kr\'olak, Kuehn, Kumar, Kumar, Kumar, Kuo, Kutynia, Kwang, Lackey, Lai, Landry, Landry, Lang, Lange, Lantz, Lanza, Lartaux-Vollard, Lasky, Laxen, Lazzarini, Lazzaro, Leaci,
  Leavey, Lee, Lee, Lee, Lee, Lee, Lehmann, Lenon, Leonardi, Leroy, Letendre, Levin, Li, Li, Li, Linker, Littenberg, Liu, Liu, Lo, Lockerbie, London, Longo, Lorenzini, Loriette, Lormand, Losurdo, Lough, Lousto, Lovelace, L\"uck, Lumaca, Lundgren, Lynch, Ma, Macas, Macfoy, Machenschalk, MacInnis, Macleod, Hernandez, Maga\~na Sandoval, Zertuche, Magee, Majorana, Maksimovic, Man, Mandic, Mangano, Mansell, Manske, Mantovani, Marchesoni, Marion, M\'arka, M\'arka, Markakis, Markosyan, Markowitz, Maros, Marquina, Martelli, Martellini, Martin, Martin, Martynov, Mason, Massera, Masserot, Massinger, Masso-Reid, Mastrogiovanni, Matas, Matichard, Matone, Mavalvala, Mazumder, McCann, McCarthy, McClelland, McCormick, McCuller, McGuire, McIver, McManus, McRae, McWilliams, Meacher, Meadors, Mehmet, Meidam, Mejuto-Villa, Melatos, Mendell, Mendoza-Gandara, Mercer, Mereni, Merilh, Merzougui, Meshkov, Messenger, Messick, Metzdorff, Meyers, Miao, Michel, Middleton, Mikhailov, Milano, Miller, Miller, Miller, Miller, Millhouse,
  Mills, Milovich-Goff, Minazzoli, Minenkov, Ming, Mishra, Mitra, Mitrofanov, Mitselmakher, Mittleman, Moffa, Mogushi, Mohan, Mohapatra, Montani, Moore, Moraru, Moreno, Morisaki, Mours, Mow-Lowry, Mueller, Muir, Mukherjee, Mukherjee, Mukherjee, Mukund, Mullavey, Munch, Mu\~niz, Muratore, Murray, Nagar, Napier, Nardecchia, Naticchioni, Nayak, Neilson, Nelemans, Nelson, Nery, Neunzert, Nevin, Newport, Ng, Ng, Nguyen, Nguyen, Nichols, Nielsen, Nissanke, Nitz, Nocera, Nolting, North, Nuttall, Obergaulinger, Oberling, O'Brien, O'Dea, Ogin, Oh, Oh, Ohme, Ohta, Okada, Oliver, Oppermann, Oram, O'Reilly, Ormiston, Ortega, O'Shaughnessy, Ossokine, Ottaway, Overmier, Owen, Pace, Pagano, Page, Page, Pai, Pai, Palamos, Palashov, Palomba, Pal-Singh, Pan, Pan, Pang, Pang, Pankow, Pannarale, Pant, Paoletti, Paoli, Papa, Parida, Parker, Pascucci, Pasqualetti, Passaquieti, Passuello, Patil, Patricelli, Pearlstone, Pedersen, Pedraza, Pedurand, Pekowsky, Pele, Penn, Perez, Perreca, Perri, Pfeiffer, Phelps, Phukon, Piccinni,
  Pichot, Piergiovanni, Pierro, Pillant, Pinard, Pinto, Pirello, Pitkin, Poggiani, Popolizio, Porter, Possenti, Post, Powell, Prasad, Pratt, Pratten, Predoi, Prestegard, Principe, Privitera, Prodi, Prokhorov, Puncken, Punturo, Puppo, P\"urrer, Qi, Quetschke, Quintero, Quitzow-James, Raab, Rabeling, Radkins, Raffai, Raja, Rajan, Rajbhandari, Rakhmanov, Ramirez, Ramos-Buades, Rana, Rapagnani, Raymond, Razzano, Read, Regimbau, Rei, Reid, Reitze, Ren, Ricci, Ricker, Riemenschneider, Riles, Rizzo, Robertson, Robie, Robinet, Robson, Rocchi, Rolland, Rollins, Roma, Romano, Romel, Romie, Rosi\ifmmode~\acute{n}\else \'{n}\fi{}ska, Ross, Rowan, R\"udiger, Ruggi, Rutins, Ryan, Sachdev, Sadecki, Sakellariadou, Salconi, Saleem, Salemi, Samajdar, Sammut, Sampson, Sanchez, Sanchez, Sanchis-Gual, Sandberg, Sanders, Sarin, Sassolas, Sathyaprakash, Saulson, Sauter, Savage, Sawadsky, Schale, Scheel, Scheuer, Schmidt, Schnabel, Schofield, Sch\"onbeck, Schreiber, Schuette, Schulte, Schutz, Schwalbe, Scott, Scott, Seidel, Sellers,
  Sengupta, Sentenac, Sequino, Sergeev, Setyawati, Shaddock, Shaffer, Shah, Shahriar, Shaner, Shao, Shapiro, Shawhan, Shen, Shoemaker, Shoemaker, Siellez, Siemens, Sieniawska, Sigg, Silva, Singer, Singh, Singhal, Sintes, Slagmolen, Slaven-Blair, Smith, Smith, Smith, Somala, Son, Sorazu, Sorrentino, Souradeep, Spencer, Srivastava, Staats, Steinke, Steinlechner, Steinlechner, Steinmeyer, Steltner, Stevenson, Stocks, Stone, Stops, Strain, Stratta, Strigin, Strunk, Sturani, Stuver, Summerscales, Sun, Sunil, Suresh, Sutton, Swinkels, Szczepa\ifmmode~\acute{n}\else \'{n}\fi{}czyk, Tacca, Tait, Talbot, Talukder, Tanner, T\'apai, Taracchini, Tasson, Taylor, Taylor, Tewari, Theeg, Thies, Thomas, Thomas, Thomas, Thorne, Thrane, Tiwari, Tiwari, Tokmakov, Toland, Tonelli, Tornasi, Torres-Forn\'e, Torrie, T\"oyr\"a, Travasso, Traylor, Trinastic, Tringali, Trozzo, Tsang, Tse, Tso, Tsuna, Tsukada, Tuyenbayev, Ueno, Ugolini, Urban, Usman, Vahlbruch, Vajente, Valdes, van Bakel, van Beuzekom, van~den Brand, Van Den~Broeck,
  Vander-Hyde, van~der Schaaf, van Heijningen, van Veggel, Vardaro, Varma, Vass, Vas\'uth, Vecchio, Vedovato, Veitch, Veitch, Venkateswara, Venugopalan, Verkindt, Vetrano, Vicer\'e, Viets, Vinciguerra, Vine, Vinet, Vitale, Vo, Vocca, Vorvick, Vyatchanin, Wade, Wade, Wade, Walet, Walker, Wallace, Walsh, Wang, Wang, Wang, Wang, Wang, Ward, Warner, Was, Watchi, Weaver, Wei, Weinert, Weinstein, Weiss, Wellmann, Wen, Wessel, We\ss{}els, Westerweck, Wette, Whelan, Whiting, Whittle, Wilken, Williams, Williams, Williamson, Willis, Willke, Wimmer, Winkler, Wipf, Wittel, Woan, Woehler, Wofford, Wong, Worden, Wright, Wu, Wysocki, Xiao, Yam, Yamamoto, Yancey, Yang, Yap, Yazback, Yu, Yu, Yvert, Zadro\ifmmode~\dot{z}\else \.{z}\fi{}ny, Zanolin, Zelenova, Zendri, Zevin, Zhang, Zhang, Zhang, Zhang, Zhang, Zhao, Zhou, Zhou, Zhu, Zhu, Zimmerman, Zlochower, Zucker, \& Zweizig}]{PhysRevX.9.011001}
---. 2019, Phys. Rev. X, 9, 011001, \dodoi{10.1103/PhysRevX.9.011001}

\bibitem[{Adhikari {et~al.}(2021{\natexlab{a}})Adhikari, Albataineh, Androic, Aniol, Armstrong, Averett, Ayerbe~Gayoso, Barcus, Bellini, Beminiwattha, Benesch, Bhatt, Bhatta~Pathak, Bhetuwal, Blaikie, Campagna, Camsonne, Cates, Chen, Clarke, Cornejo, Covrig~Dusa, Datta, Deshpande, Dutta, Feldman, Fuchey, Gal, Gaskell, Gautam, Gericke, Ghosh, Halilovic, Hansen, Hauenstein, Henry, Horowitz, Jantzi, Jian, Johnston, Jones, Karki, Katugampola, Keppel, King, King, Knauss, Kumar, Kutz, Lashley-Colthirst, Leverick, Liu, Liyange, Malace, Mammei, Mammei, McCaughan, McNulty, Meekins, Metts, Michaels, Mondal, Napolitano, Narayan, Nikolaev, Rashad, Owen, Palatchi, Pan, Pandey, Park, Paschke, Petrusky, Pitt, Premathilake, Puckett, Quinn, Radloff, Rahman, Rathnayake, Reed, Reimer, Richards, Riordan, Roblin, Seeds, Shahinyan, Souder, Tang, Thiel, Tian, Urciuoli, Wertz, Wojtsekhowski, Yale, Ye, Yoon, Zec, Zhang, Zhang, \& Zheng}]{prex}
Adhikari, D., Albataineh, H., Androic, D., {et~al.} 2021{\natexlab{a}}, Phys. Rev. Lett., 126, 172502, \dodoi{10.1103/PhysRevLett.126.172502}

\bibitem[{Adhikari {et~al.}(2021{\natexlab{b}})}]{PREX:2021umo}
Adhikari, D., {et~al.} 2021{\natexlab{b}}, Phys. Rev. Lett., 126, 172502, \dodoi{10.1103/PhysRevLett.126.172502}

\bibitem[{Adhikari {et~al.}(2022)Adhikari, Albataineh, Androic, Aniol, Armstrong, Averett, Ayerbe~Gayoso, Barcus, Bellini, Beminiwattha, Benesch, Bhatt, Bhatta~Pathak, Bhetuwal, Blaikie, Boyd, Campagna, Camsonne, Cates, Chen, Clarke, Cornejo, Covrig~Dusa, Dalton, Datta, Deshpande, Dutta, Feldman, Fuchey, Gal, Gaskell, Gautam, Gericke, Ghosh, Halilovic, Hansen, Hassan, Hauenstein, Henry, Horowitz, Jantzi, Jian, Johnston, Jones, Kakkar, Katugampola, Keppel, King, King, Kumar, Kutz, Lashley-Colthirst, Leverick, Liu, Liyanage, Mammei, Mammei, McCaughan, McNulty, Meekins, Metts, Michaels, Mihovilovic, Mondal, Napolitano, Narayan, Nikolaev, Owen, Palatchi, Pan, Pandey, Park, Paschke, Petrusky, Pitt, Premathilake, Quinn, Radloff, Rahman, Rashad, Rathnayake, Reed, Reimer, Richards, Riordan, Roblin, Seeds, Shahinyan, Souder, Thiel, Tian, Urciuoli, Wertz, Wojtsekhowski, Yale, Ye, Yoon, Xiong, Zec, Zhang, Zhang, \& Zheng}]{crex}
Adhikari, D., Albataineh, H., Androic, D., {et~al.} 2022, Phys. Rev. Lett., 129, 042501, \dodoi{10.1103/PhysRevLett.129.042501}

\bibitem[{Annala {et~al.}(2018)Annala, Gorda, Kurkela, \& Vuorinen}]{Eemeli}
Annala, E., Gorda, T., Kurkela, A., \& Vuorinen, A. 2018, Physical Review Letters, 120, \dodoi{10.1103/physrevlett.120.172703}

\bibitem[{{Barret} {et~al.}(2013){Barret}, {Nandra}, {Barcons}, {Fabian}, {den Herder}, {Piro}, {Watson}, {Aird}, {Branduardi-Raymont}, {Cappi}, {Carrera}, {Comastri}, {Costantini}, {Croston}, {Decourchelle}, {Done}, {Dovciak}, {Ettori}, {Finoguenov}, {Georgakakis}, {Jonker}, {Kaastra}, {Matt}, {Motch}, {O'Brien}, {Pareschi}, {Pointecouteau}, {Pratt}, {Rauw}, {Reiprich}, {Sanders}, {Sciortino}, {Willingale}, \& {Wilms}}]{2013sf2a.conf..447B}
{Barret}, D., {Nandra}, K., {Barcons}, X., {et~al.} 2013, in SF2A-2013: Proceedings of the Annual meeting of the French Society of Astronomy and Astrophysics, ed. L.~{Cambresy}, F.~{Martins}, E.~{Nuss}, \& A.~{Palacios}, 447--453, \dodoi{10.48550/arXiv.1310.3814}

\bibitem[{{Bogdanov} {et~al.}(2019){Bogdanov}, {Lamb}, {Mahmoodifar}, {Miller}, {Morsink}, {Riley}, {Strohmayer}, {Tung}, {Watts}, {Dittmann}, {Chakrabarty}, {Guillot}, {Arzoumanian}, \& {Gendreau}}]{2019ApJ...887L..26B}
{Bogdanov}, S., {Lamb}, F.~K., {Mahmoodifar}, S., {et~al.} 2019, \apjl, 887, L26, \dodoi{10.3847/2041-8213/ab5968}

\bibitem[{{Buchner}(2021)}]{2021JOSS....6.3001B}
{Buchner}, J. 2021, The Journal of Open Source Software, 6, 3001, \dodoi{10.21105/joss.03001}

\bibitem[{Choudhury {et~al.}(2024)Choudhury, Salmi, Vinciguerra, Riley, Kini, Watts, Dorsman, Bogdanov, Guillot, Ray, Reardon, Remillard, Bilous, Huppenkothen, Lattimer, Rutherford, Arzoumanian, Gendreau, Morsink, \& Ho}]{Choudhury_2024}
Choudhury, D., Salmi, T., Vinciguerra, S., {et~al.} 2024, The Astrophysical Journal Letters, 971, L20, \dodoi{10.3847/2041-8213/ad5a6f}

\bibitem[{{Dittmann} {et~al.}(2024){Dittmann}, {Miller}, {Lamb}, {Holt}, {Chirenti}, {Wolff}, {Bogdanov}, {Guillot}, {Ho}, {Morsink}, {Arzoumanian}, \& {Gendreau}}]{2024ApJ...974..295D}
{Dittmann}, A.~J., {Miller}, M.~C., {Lamb}, F.~K., {et~al.} 2024, \apj, 974, 295, \dodoi{10.3847/1538-4357/ad5f1e}

\bibitem[{{Drischler} {et~al.}(2021){Drischler}, {Han}, {Lattimer}, {Prakash}, {Reddy}, \& {Zhao}}]{Drischler21}
{Drischler}, C., {Han}, S., {Lattimer}, J.~M., {et~al.} 2021, \prc, 103, 045808, \dodoi{10.1103/PhysRevC.103.045808}

\bibitem[{Essick {et~al.}(2021)Essick, Tews, Landry, \& Schwenk}]{Essick2021}
Essick, R., Tews, I., Landry, P., \& Schwenk, A. 2021, \prl, 127, 192701, \dodoi{10.1103/PhysRevLett.127.192701}

\bibitem[{{Fonseca} {et~al.}(2021){Fonseca}, {Cromartie}, {Pennucci}, {Ray}, {Kirichenko}, {Ransom}, {Demorest}, {Stairs}, {Arzoumanian}, {Guillemot}, {Parthasarathy}, {Kerr}, {Cognard}, {Baker}, {Blumer}, {Brook}, {DeCesar}, {Dolch}, {Dong}, {Ferrara}, {Fiore}, {Garver-Daniels}, {Good}, {Jennings}, {Jones}, {Kaspi}, {Lam}, {Lorimer}, {Luo}, {McEwen}, {McKee}, {McLaughlin}, {McMann}, {Meyers}, {Naidu}, {Ng}, {Nice}, {Pol}, {Radovan}, {Shapiro-Albert}, {Tan}, {Tendulkar}, {Swiggum}, {Wahl}, \& {Zhu}}]{Fonseca21}
{Fonseca}, E., {Cromartie}, H.~T., {Pennucci}, T.~T., {et~al.} 2021, \apjl, 915, L12, \dodoi{10.3847/2041-8213/ac03b8}

\bibitem[{Froning {et~al.}(2024)Froning, Roming, Ray, Argan, Arzoumanian, Bogdanov, Bonvicini, Brandt, Bursa, Cackett, {et~al.}}]{froning2024strobe}
Froning, C.~S., Roming, P., Ray, P., {et~al.} 2024, in Space Telescopes and Instrumentation 2024: Ultraviolet to Gamma Ray, Vol. 13093, SPIE, 729--735

\bibitem[{Gendreau {et~al.}(2016)Gendreau, Arzoumanian, Adkins, Albert, Anders, Aylward, Baker, Balsamo, Bamford, Benegalrao, {et~al.}}]{gendreau2016neutron}
Gendreau, K.~C., Arzoumanian, Z., Adkins, P.~W., {et~al.} 2016, in Space telescopes and instrumentation 2016: Ultraviolet to gamma ray, Vol. 9905, SPIE, 420--435

\bibitem[{Huang(2024)}]{Huang}
Huang, C. 2024, in prep

\bibitem[{Huang \& Chen(2024)}]{chun24hotspot}
Huang, C., \& Chen, A. 2024, in prep

\bibitem[{Huang {et~al.}(2024{\natexlab{a}})Huang, Raaijmakers, Watts, Tolos, \& Provid\^encia}]{Huang:2023grj}
Huang, C., Raaijmakers, G., Watts, A.~L., Tolos, L., \& Provid\^encia, C. 2024{\natexlab{a}}, Mon. Not. Roy. Astron. Soc., 529, 4650, \dodoi{10.1093/mnras/stae844}

\bibitem[{Huang {et~al.}(2024{\natexlab{b}})Huang, Tolos, Provid\^encia, \& Watts}]{Huang:2024rvj}
Huang, C., Tolos, L., Provid\^encia, C., \& Watts, A. 2024{\natexlab{b}}.
\newblock \doarXiv{2410.14572}

\bibitem[{Huang {et~al.}(2024{\natexlab{c}})Huang, Malik, Cartaxo, Sourav, Yuan, Zhou, Liu, Groger, Dong, Osborn, Whitsett, Wang, Provid\^{e}ncia, Oertel, Chen, Tolos, \& Watts}]{compactobject}
Huang, C., Malik, T., Cartaxo, J., {et~al.} 2024{\natexlab{c}}, {Journal of Open Source Software}.
\newblock \doarXiv{2411.14615}

\bibitem[{Kumar {et~al.}(2023)Kumar, Kumar, Thakur, Kumar, Agrawal, \& Dhiman}]{kumar23}
Kumar, M., Kumar, S., Thakur, V., {et~al.} 2023, Phys. Rev. C, 107, 055801, \dodoi{10.1103/PhysRevC.107.055801}

\bibitem[{{Lattimer} \& {Lim}(2013)}]{Lattimer2013}
{Lattimer}, J.~M., \& {Lim}, Y. 2013, \apj, 771, 51, \dodoi{10.1088/0004-637X/771/1/51}

\bibitem[{{Lattimer} \& {Prakash}(2001)}]{Lattimer2001}
{Lattimer}, J.~M., \& {Prakash}, M. 2001, \apj, 550, 426, \dodoi{10.1086/319702}

\bibitem[{{Lim} \& {Schwenk}(2024)}]{lim24}
{Lim}, Y., \& {Schwenk}, A. 2024, \prc, 109, 035801, \dodoi{10.1103/PhysRevC.109.035801}

\bibitem[{Miller {et~al.}(2019)Miller, Lamb, Dittmann, Bogdanov, Arzoumanian, Gendreau, Guillot, Harding, Ho, Lattimer, Ludlam, Mahmoodifar, Morsink, Ray, Strohmayer, Wood, Enoto, Foster, Okajima, Prigozhin, \& Soong}]{Miller_2019}
Miller, M.~C., Lamb, F.~K., Dittmann, A.~J., {et~al.} 2019, The Astrophysical Journal Letters, 887, L24, \dodoi{10.3847/2041-8213/ab50c5}

\bibitem[{{Miller} {et~al.}(2021){Miller}, {Lamb}, {Dittmann}, {Bogdanov}, {Arzoumanian}, {Gendreau}, {Guillot}, {Ho}, {Lattimer}, {Loewenstein}, {Morsink}, {Ray}, {Wolff}, {Baker}, {Cazeau}, {Manthripragada}, {Markwardt}, {Okajima}, {Pollard}, {Cognard}, {Cromartie}, {Fonseca}, {Guillemot}, {Kerr}, {Parthasarathy}, {Pennucci}, {Ransom}, \& {Stairs}}]{Miller21}
{Miller}, M.~C., {Lamb}, F.~K., {Dittmann}, A.~J., {et~al.} 2021, \apjl, 918, L28, \dodoi{10.3847/2041-8213/ac089b}

\bibitem[{Miller {et~al.}(2021)Miller, Lamb, Dittmann, Bogdanov, Arzoumanian, Gendreau, Guillot, Ho, Lattimer, Loewenstein, Morsink, Ray, Wolff, Baker, Cazeau, Manthripragada, Markwardt, Okajima, Pollard, Cognard, Cromartie, Fonseca, Guillemot, Kerr, Parthasarathy, Pennucci, Ransom, \& Stairs}]{Miller_2021}
Miller, M.~C., Lamb, F.~K., Dittmann, A.~J., {et~al.} 2021, The Astrophysical Journal Letters, 918, L28, \dodoi{10.3847/2041-8213/ac089b}

\bibitem[{Mondal \& Gulminelli(2023)}]{Mondal:2022cva}
Mondal, C., \& Gulminelli, F. 2023, Phys. Rev. C, 107, 015801, \dodoi{10.1103/PhysRevC.107.015801}

\bibitem[{Provid\^encia \& Rabhi(2013)}]{PhysRevC.87.055801}
Provid\^encia, C. m.~c., \& Rabhi, A. 2013, Phys. Rev. C, 87, 055801, \dodoi{10.1103/PhysRevC.87.055801}

\bibitem[{Raaijmakers {et~al.}(2019)Raaijmakers, Riley, Watts, Greif, Morsink, Hebeler, Schwenk, Hinderer, Nissanke, Guillot, Arzoumanian, Bogdanov, Chakrabarty, Gendreau, Ho, Lattimer, Ludlam, \& Wolff}]{Raaijmakers_2019}
Raaijmakers, G., Riley, T.~E., Watts, A.~L., {et~al.} 2019, The Astrophysical Journal Letters, 887, L22, \dodoi{10.3847/2041-8213/ab451a}

\bibitem[{Raaijmakers {et~al.}(2020)Raaijmakers, Greif, Riley, Hinderer, Hebeler, Schwenk, Watts, Nissanke, Guillot, Lattimer, \& Ludlam}]{Raaijmakers_2020}
Raaijmakers, G., Greif, S.~K., Riley, T.~E., {et~al.} 2020, The Astrophysical Journal Letters, 893, L21, \dodoi{10.3847/2041-8213/ab822f}

\bibitem[{Raaijmakers {et~al.}(2021)Raaijmakers, Greif, Hebeler, Hinderer, Nissanke, Schwenk, Riley, Watts, Lattimer, \& Ho}]{Raaijmakers_2021}
Raaijmakers, G., Greif, S.~K., Hebeler, K., {et~al.} 2021, The Astrophysical Journal Letters, 918, L29, \dodoi{10.3847/2041-8213/ac089a}

\bibitem[{{Ray} {et~al.}(2019){Ray}, {Arzoumanian}, {Ballantyne}, {Bozzo}, {Brandt}, {Brenneman}, {Chakrabarty}, {Christophersen}, {DeRosa}, {Feroci}, {Gendreau}, {Goldstein}, {Hartmann}, {Hernanz}, {Jenke}, {Kara}, {Maccarone}, {McDonald}, {Nowak}, {Phlips}, {Remillard}, {Stevens}, {Tomsick}, {Watts}, {Wilson-Hodge}, {Wood}, {Zane}, {Ajello}, {Alston}, {Altamirano}, {Antoniou}, {Arur}, {Ashton}, {Auchettl}, {Ayres}, {Bachetti}, {Balokovic}, {Baring}, {Baykal}, {Begelman}, {Bhat}, {Bogdanov}, {Briggs}, {Bulbul}, {Bult}, {Burns}, {Cackett}, {Campana}, {Caspi}, {Cavecchi}, {Chenevez}, {Cherry}, {Corbet}, {Corcoran}, {Corsi}, {Degenaar}, {Drake}, {Eikenberry}, {Enoto}, {Fragile}, {Fuerst}, {Gandhi}, {Garcia}, {Goldstein}, {Gonzalez}, {Grefenstette}, {Grinberg}, {Grossan}, {Guillot}, {Guver}, {Haggard}, {Heinke}, {Heinz}, {Hemphill}, {Homan}, {Hui}, {Huppenkothen}, {Ingram}, {Irwin}, {Jaisawal}, {Jaodand}, {Kalemci}, {Kaplan}, {Keek}, {Kennea}, {Kerr}, {van der Klis}, {Kocevski}, {Koss}, {Kowalski}, {Lai},
  {Lamb}, {Laycock}, {Lazio}, {Lazzati}, {Longcope}, {Loewenstein}, {Maitra}, {Majid}, {Maksym}, {Malacaria}, {Margutti}, {Martindale}, {McHardy}, {Meyer}, {Middleton}, {Miller}, {Miller}, {Motta}, {Neilsen}, {Nelson}, {Noble}, {O'Brien}, {Osborne}, {Osten}, {Ozel}, {Palliyaguru}, {Pasham}, {Patruno}, {Pelassa}, {Petropoulou}, {Pilia}, {Pohl}, {Pooley}, {Prescod-Weinstein}, {Psaltis}, {Raaijmakers}, {Reynolds}, {Riley}, {Salvesen}, {Santangelo}, {Scaringi}, {Schanne}, {Schnittman}, {Smith}, {Smith}, {Snios}, {Steiner}, {Steiner}, {Stella}, {Strohmayer}, {Sun}, {Tauris}, {Taylor}, {Tohuvavohu}, {Vacchi}, {Vasilopoulos}, {Veledina}, {Walsh}, {Weinberg}, {Wilkins}, {Willingale}, {Wilms}, {Winter}, {Wolff}, {in 't Zand}, {Zezas}, {Zhang}, \& {Zoghbi}}]{strobex}
{Ray}, P.~S., {Arzoumanian}, Z., {Ballantyne}, D., {et~al.} 2019, arXiv e-prints, arXiv:1903.03035.
\newblock \doarXiv{1903.03035}

\bibitem[{Reardon {et~al.}(2024)Reardon, Bailes, Shannon, Flynn, Askew, Bhat, Chen, Curyło, Feng, Hobbs, Kapur, Kerr, Liu, Manchester, Mandow, Mishra, Russell, Shamohammadi, Zhang, \& Zic}]{Reardon_2024}
Reardon, D.~J., Bailes, M., Shannon, R.~M., {et~al.} 2024, The Astrophysical Journal Letters, 971, L18, \dodoi{10.3847/2041-8213/ad614a}

\bibitem[{Reed {et~al.}(2021)Reed, Fattoyev, Horowitz, \& Piekarewicz}]{Reed_2021}
Reed, B.~T., Fattoyev, F., Horowitz, C., \& Piekarewicz, J. 2021, Physical Review Letters, 126, \dodoi{10.1103/physrevlett.126.172503}

\bibitem[{Reinhard {et~al.}(2013)Reinhard, Piekarewicz, Nazarewicz, Agrawal, Paar, \& Roca-Maza}]{Reinhard13}
Reinhard, P.-G., Piekarewicz, J., Nazarewicz, W., {et~al.} 2013, Phys. Rev. C, 88, 034325, \dodoi{10.1103/PhysRevC.88.034325}

\bibitem[{Reinhard {et~al.}(2021)Reinhard, Roca-Maza, \& Nazarewicz}]{Reinhard21}
Reinhard, P.-G., Roca-Maza, X., \& Nazarewicz, W. 2021, Phys. Rev. Lett., 127, 232501, \dodoi{10.1103/PhysRevLett.127.232501}

\bibitem[{Reinhard {et~al.}(2022)Reinhard, Roca-Maza, \& Nazarewicz}]{Reinhard:2022inh}
---. 2022, Phys. Rev. Lett., 129, 232501, \dodoi{10.1103/PhysRevLett.129.232501}

\bibitem[{Riley {et~al.}(2019)Riley, Watts, Bogdanov, Ray, Ludlam, Guillot, Arzoumanian, Baker, Bilous, Chakrabarty, Gendreau, Harding, Ho, Lattimer, Morsink, \& Strohmayer}]{Riley_2019}
Riley, T.~E., Watts, A.~L., Bogdanov, S., {et~al.} 2019, The Astrophysical Journal Letters, 887, L21, \dodoi{10.3847/2041-8213/ab481c}

\bibitem[{{Riley} {et~al.}(2021){Riley}, {Watts}, {Ray}, {Bogdanov}, {Guillot}, {Morsink}, {Bilous}, {Arzoumanian}, {Choudhury}, {Deneva}, {Gendreau}, {Harding}, {Ho}, {Lattimer}, {Loewenstein}, {Ludlam}, {Markwardt}, {Okajima}, {Prescod-Weinstein}, {Remillard}, {Wolff}, {Fonseca}, {Cromartie}, {Kerr}, {Pennucci}, {Parthasarathy}, {Ransom}, {Stairs}, {Guillemot}, \& {Cognard}}]{Riley21}
{Riley}, T.~E., {Watts}, A.~L., {Ray}, P.~S., {et~al.} 2021, \apjl, 918, L27, \dodoi{10.3847/2041-8213/ac0a81}

\bibitem[{Rutherford {et~al.}(2024)Rutherford, Mendes, Svensson, Schwenk, Watts, Hebeler, Keller, Prescod-Weinstein, Choudhury, Raaijmakers, Salmi, Timmerman, Vinciguerra, Guillot, \& Lattimer}]{Rutherford_2024}
Rutherford, N., Mendes, M., Svensson, I., {et~al.} 2024, The Astrophysical Journal Letters, 971, L19, \dodoi{10.3847/2041-8213/ad5f02}

\bibitem[{{Salmi} {et~al.}(2022){Salmi}, {Vinciguerra}, {Choudhury}, {Riley}, {Watts}, {Remillard}, {Ray}, {Bogdanov}, {Guillot}, {Arzoumanian}, {Chirenti}, {Dittmann}, {Gendreau}, {Ho}, {Miller}, {Morsink}, {Wadiasingh}, \& {Wolff}}]{Salmi2022}
{Salmi}, T., {Vinciguerra}, S., {Choudhury}, D., {et~al.} 2022, \apj, 941, 150, \dodoi{10.3847/1538-4357/ac983d}

\bibitem[{{Salmi} {et~al.}(2024){Salmi}, {Choudhury}, {Kini}, {Riley}, {Vinciguerra}, {Watts}, {Wolff}, {Arzoumanian}, {Bogdanov}, {Chakrabarty}, {Gendreau}, {Guillot}, {Ho}, {Huppenkothen}, {Ludlam}, {Morsink}, \& {Ray}}]{salmi2024}
{Salmi}, T., {Choudhury}, D., {Kini}, Y., {et~al.} 2024, arXiv e-prints, arXiv:2406.14466, \dodoi{10.48550/arXiv.2406.14466}

\bibitem[{Salmi {et~al.}(2024)Salmi, Choudhury, Kini, Riley, Vinciguerra, Watts, Wolff, Arzoumanian, Bogdanov, Chakrabarty, Gendreau, Guillot, Ho, Huppenkothen, Ludlam, Morsink, \& Ray}]{salmi2024radiushighmasspulsar}
Salmi, T., Choudhury, D., Kini, Y., {et~al.} 2024, The Radius of the High Mass Pulsar PSR J0740+6620 With 3.6 Years of NICER Data.
\newblock \doarXiv{2406.14466}

\bibitem[{Scurto {et~al.}(2023)Scurto, Pais, \& Gulminelli}]{Scurto_2023}
Scurto, L., Pais, H., \& Gulminelli, F. 2023, Physical Review C, 107, \dodoi{10.1103/physrevc.107.045806}

\bibitem[{Vinciguerra {et~al.}(2024)Vinciguerra, Salmi, Watts, Choudhury, Riley, Ray, Bogdanov, Kini, Guillot, Chakrabarty, Ho, Huppenkothen, Morsink, Wadiasingh, \& Wolff}]{Vinciguerra_2024}
Vinciguerra, S., Salmi, T., Watts, A.~L., {et~al.} 2024, The Astrophysical Journal, 961, 62, \dodoi{10.3847/1538-4357/acfb83}

\bibitem[{{Watts} {et~al.}(2016){Watts}, {Andersson}, {Chakrabarty}, {Feroci}, {Hebeler}, {Israel}, {Lamb}, {Miller}, {Morsink}, {{\"O}zel}, {Patruno}, {Poutanen}, {Psaltis}, {Schwenk}, {Steiner}, {Stella}, {Tolos}, \& {van der Klis}}]{Watts_2016}
{Watts}, A.~L., {Andersson}, N., {Chakrabarty}, D., {et~al.} 2016, Reviews of Modern Physics, 88, 021001, \dodoi{10.1103/RevModPhys.88.021001}

\bibitem[{{Watts} {et~al.}(2019){Watts}, {Yu}, {Poutanen}, {Zhang}, {Bhattacharyya}, {Bogdanov}, {Ji}, {Patruno}, {Riley}, {Bakala}, {Baykal}, {Bernardini}, {Bombaci}, {Brown}, {Cavecchi}, {Chakrabarty}, {Chenevez}, {Degenaar}, {Del Santo}, {Di Salvo}, {Doroshenko}, {Falanga}, {Ferdman}, {Feroci}, {Gambino}, {Ge}, {Greif}, {Guillot}, {Gungor}, {Hartmann}, {Hebeler}, {Heger}, {Homan}, {Iaria}, {Zand}, {Kargaltsev}, {Kurkela}, {Lai}, {Li}, {Li}, {Li}, {Linares}, {Lu}, {Mahmoodifar}, {M{\'e}ndez}, {Coleman Miller}, {Morsink}, {N{\"a}ttil{\"a}}, {Possenti}, {Prescod-Weinstein}, {Qu}, {Riggio}, {Salmi}, {Sanna}, {Santangelo}, {Schatz}, {Schwenk}, {Song}, {{\v{S}}r{\'a}mkov{\'a}}, {Stappers}, {Stiele}, {Strohmayer}, {Tews}, {Tolos}, {T{\"o}r{\"o}k}, {Tsang}, {Urbanec}, {Vacchi}, {Xu}, {Xu}, {Zane}, {Zhang}, {Zhang}, {Zhang}, {Zheng}, \& {Zhou}}]{extp_watts}
{Watts}, A.~L., {Yu}, W., {Poutanen}, J., {et~al.} 2019, Science China Physics, Mechanics, and Astronomy, 62, 29503, \dodoi{10.1007/s11433-017-9188-4}

\bibitem[{Yue {et~al.}(2022)Yue, Chen, Zhang, \& Zhou}]{Yue:2021yfx}
Yue, T.-G., Chen, L.-W., Zhang, Z., \& Zhou, Y. 2022, Phys. Rev. Res., 4, L022054, \dodoi{10.1103/PhysRevResearch.4.L022054}

\end{thebibliography}
\bibliographystyle{aasjournal}



\end{document}